\documentclass{aa}  

\usepackage{xcolor}

\usepackage{graphicx}

\usepackage{txfonts}

\usepackage{ulem}

\usepackage{placeins}

\begin{document} 

   \title{The soft X-ray transient EP241021A: A cosmic explosion with a complex off-axis jet and cocoon from a massive progenitor}

   \titlerunning{EP241021a: a soft GRB from a massive star progenitor}

\author{
Giulia Gianfagna\inst{1}
\and Luigi Piro\inst{1}
\and Gabriele Bruni\inst{1}
\and Aishwarya Linesh Thakur\inst{1}
\and Hendrik Van Eerten\inst{2}
\and Maria D. Caballero-García\inst{3}
\and Alberto Castro-Tirado\inst{3}
\and Yong Chen\inst{4}
\and Ye-hao Cheng\inst{5}
\and Maria Gritsevich\inst{6,7}
\and Sergiy Guziy\inst{3}
\and Han He\inst{8}
\and You-Dong Hu\inst{9}
\and Shumei Jia\inst{4}
\and Zhixing Ling\inst{10,11}
\and Elisabetta Maiorano\inst{12}
\and Rosita Paladino\inst{13}
\and Shashi B. Pandey\inst{14}
\and Roberta Tripodi\inst{15}
\and Andrea Rossi\inst{12}
\and Rubén Sánchez-Ramírez\inst{3}
\and Shuaikang Yang\inst{8}
\and Jianghui Yuan\inst{2}
\and Weimin Yuan\inst{10,11}
\and Chen Zhang\inst{10}
}

\institute{
INAF -- Istituto di Astrofisica e Planetologia Spaziali, via Fosso del Cavaliere 100, I-00133 Rome, Italy\\
\email{giulia.gianfagna@inaf.it}
\and
Department of Physics, University of Bath, Claverton Down, Bath, BA2 7AY, UK
\and
Instituto de Astrofísica de Andalucía (IAA-CSIC), Glorieta de la Astronomía s/n, 18008, Granada, Spain
\and
Key Laboratory of Particle Astrophysics, Institute of High Energy Physics, Chinese Academy of Sciences, Beijing 100049, China
\and
South-Western Institute for Astronomy Research, Yunnan University, Kunming, Yunnan 650500, China
\and
Faculty of Science, University of Helsinki, Gustaf Hällströmin katu 2a, P.O. Box 64, FI-00014 Helsinki, Finland
\and
Institute of Physics and Technology, Ural Federal University, Mira str. 19, 620002 Ekaterinburg, Russia
\and
Department of Astronomy, School of Physics and Technology, Wuhan University, Wuhan 430072, China
\and
Guangxi Key Laboratory for Relativistic Astrophysics, School of Physical Science and Technology, Guangxi University, Nanning 530004, China
\and
National Astronomical Observatories, Chinese Academy of Sciences, Beijing 100101, China
\and
School of Astronomy and Space Science, University of Chinese Academy of Sciences, Chinese Academy of Sciences, Beijing 100049, China
\and
INAF -- Osservatorio di Astrofisica e Scienza dello Spazio, via Piero Gobetti 93/3, I-40129 Bologna, Italy
\and
INAF -- Istituto di Radioastronomia, Via P. Gobetti 101, 40129 Bologna, Italy
\and
ARIES -- Aryabhatta Research Institute of Observational Sciences, Nainital, 263001, India
\and
INAF -- Osservatorio Astronomico di Roma, Via Frascati 33, Monte Porzio Catone, 00078, Italy
}

   \date{Received , ; accepted , }

\abstract
   {X-ray flashes (XRFs) are fast X-ray transients discovered by the \textit{BeppoSAX} satellite.
   Diverse evidence indicates that XRFs are connected to gamma ray bursts (GRBs) and likely represent their softer analogs. With its soft X-ray bandpass and exquisite sensitivity, the \textit{Einstein Probe} (EP) offers a novel opportunity to disclose the nature of such puzzling events.}
   {Several models have been proposed to explain the observed properties of XRFs, mostly in the context of the collapsar scenario, where such soft events could have different geometrical or physical conditions of the progenitor with respect to GRBs. These include off-axis GRBs and baryon-loaded explosions, which either produce a low-Lorentz-factor jet or a spherical, mildly (or non-) relativistic ejecta, known as cocoons. In this paper, we present multiwavelength observations of the afterglow of EP241021a, a soft X-ray transient detected by EP. We attempt to connect the complex, multicomponent afterglow emission with leading XRF models.}
   {We first characterize the prompt emission of EP2410121a by EP-WXT and Fermi-GBM. Then, we present the results of our multiwavelength campaign from radio (uGMRT, ATCA, e-MERLIN, and ALMA), optical (LBT, GTC, and CAHA) and X-rays (EP-FXT). We perform an analysis of light curves and broad-band spectra using both empirical and physical models of GRBs and spherical expansions (both nonrelativistic and mildly relativistic cocoons).}
   {The EP241021a afterglow is characterized by multiple components, which represent the imprints of the interaction of a jet with the complex environment of the preexisting progenitor that is likely shaping its structure. In particular, the optical and X-ray afterglows are well described by a structured jet with wide and low-Lorentz-factor ($\gamma\sim40$) wings, which produce the decreasing light curve before 6 days. A re-brightening at 7 days in the optical and X-ray data is due to the jet core, which is off-axis and coming into view. The radio emission can be modeled with a mildly relativistic cocoon ($\gamma\sim2$). Finally, in the radio spectrum at 70 days, we find an additional component peaking at $\sim50$ GHz, which is well described by a second cocoon with $\gamma\sim1$}
   {}

   \maketitle

\section{Introduction}

The key properties of long gamma ray bursts (GRBs) are successfully explained by the so-called "collapsar" scenario. After the core-collapse explosion of a stripped-envelope massive star -- likely a rapidly rotating Wolf-Rayet (WR, \citealt{Yoon2005}) -- matter flows toward a newly formed black hole or rapidly spinning, highly magnetized neutron star \citep{Woosley1993, MacFadyen1999}, and a relativistic jet breaks free from the stellar progenitor along the polar axis \citep{MacFadyen1999, Aloy2000}. Internal shocks produce the gamma ray (prompt emission) and, when the ejecta reach the circumstellar medium, they emit the afterglow (detected in X-ray, optical, and radio bands). The observer can be within the jet cone angle (on-axis) or outside it (off-axis). When on-axis, prompt emission in the gamma rays is observed; instead, when viewed off-axis, either no prompt emission or some less energetic emission from the tails of the jet may be observed.

However, the collapsar scenario also predicts the presence of additional components -- jetted or spherical \citep{Gottlieb2020, Salafia2020, RamirezRuiz2002, MacFadyen2001} -- which can manifest depending on various conditions. 
Before the relativistic jet exits the progenitor star, part of the energy output is deposited into a “cocoon” surrounding it. As the jet head breaks out of the progenitor star, the cocoon plasma can escape swiftly from the stellar cavity and accelerate.
Instead, as the cocoon itself breaks out of the stellar envelope, prompt emission in the soft keV energies may be seen prior to the typical gamma-ray emission of GRBs. These precursors have already been observed in a small number of GRBs (recently in \citealt{EP240315a_liu2024}; but see also \citealt{Piro2005, Lazzati2005}). 

There is also the possibility that the jet is polluted by baryons (the so-called baryon-loaded jet). While long-duration GRB afterglows are powered by explosions that have minimal initial ejecta mass, typically $10^{-6} M_{\odot}$, in the baryon-loaded case the jet has a mass of $M\sim 10^{-5} - 10^{-3} \rm M_{\odot}$. In this case, the jet does not reach the high Lorentz factors ($10 < \gamma_0 < 100$) necessary to produce a prompt emission in the gamma rays, and the prompt emission is expected in the soft X-rays. The expected initial energies are similar to GRBs, $10^{51} - 10^{53}$ erg \citep{MacFadyen2001, Huang2004}.  
If baryon loading is high, the jet can also be choked entirely (choked-jet scenario), but a small amount of spherical and relativistic ejecta is still launched \citep{Chakraborti2011}. The observational signature in this scenario is either a relativistic supernova (SN, as SN 2012ap with a mass $>10^{-2.5}M_{\odot}$, \citealt{Chakraborti2015, Margutti2014, Soderberg2010}), or a classical nonrelativistic SN (with a mass of 0.1 $M_{\odot}$). 

Observations of GRBs in the context of the collapsar model typically reveal only the presence of a relativistic jet. In fact, there is only limited evidence suggesting that baryon-loaded explosions can produce low-luminosity GRBs, such as GRB 060218 \citep{Campana2006, Soderberg2006}, GRB 100316D \citep{Starling2011, Margutti2013}, GRB 980425 \citep{Pian1999}, and GRB 031203 \citep{Watson2004}. This is likely attributable to the dominance of relativistic jet emission over other potential components, but it begs the question whether such other components, expected in the collapsar scenario, can shape other classes of cosmic explosions.
X-ray flashes (XRFs) are a very promising candidate. They are fast X-ray transients discovered by the \textit{BeppoSAX} satellite \citep{Heise2003}, classified as GRBs with absent (or very low) gamma-ray emission. Including the somewhat harder X-ray-rich (XRR) GRBs, they make up about two-thirds of GRB events observed by \textit{BeppoSAX} or \textit{HETE2} \citep[for a review]{Piro12}. This interesting class of events can potentially bridge the gap between highly collimated and relativistic jets and baryon-loaded explosions and cocoons.

The XRFs show an isotropic distribution on the sky and a duration of the prompt emission between 10-1000 seconds.
The observed prompt X-ray properties are similar to those of GRBs, except for the softer peak energy of the spectrum \citep{Barraud2003, DAlessio_2006}. For example, several follow the Amati - Yonetoku relation  \citep{Willingale2017}. Several pieces of evidence (including prompt phase observations by \textit{BeppoSAX} and \textit{HETE-2}; see \citealt{Sakamoto_2008}) indicate that these transients are connected to GRBs and most likely represent their softer analogs.
They lie at extragalactic distances, with host galaxies not unlike those of long GRBs \citep{Chen2021, DAlessio_2006}, and they often have similar X-ray to radio afterglow emission \citep{Bi2018, ChandraFrail2012, Sakamoto_2008, Gendre2007, DAlessio_2006}. Many XRFs are also associated with SNe \citep{Bi2018}, which further supports that they share the same origin as GRBs.
 
Even if they are likely associated with the collapsar scenario, the XRF emission origin is still unclear. Several models have been proposed to explain the observed properties in the context of the collapsar scenario depicted above. The main ones are:

(1) Off-axis GRBs \citep{Zhang2003, Yamazaki2002, Granot2002, Granot2005, Urata2015}. In this case, the observer's viewing angle (the angle between the line of sight and the jet axis) is outside the jet's cone. The narrow, collimated relativistic jet (emitting gamma rays) is not seen, but the jet tail emission at lower Lorentz factors can still be detected in X-rays. From their afterglows, however, there is no compelling evidence of XRFs coming from off-axis jets \citep{DAlessio_2006}.

(2) Baryon-loaded or choked jets, also called sub-energetic or inefficient fireballs \citep{Huang2002, Rhoads2003, Dermer2000, Zhang2002}, which also lead to the cocoon emission mentioned before. In this case, the ejecta are less energetic than a GRB jet and therefore emit the prompt emission in the X-rays.

(3) GRBs at such high redshifts that the observed peak is moved into the X-ray band \citep{Heise2003}. This, however, can be excluded as the only source of XRFs, as this should also result in fainter afterglows, which are not observed \citep{DAlessio_2006}. Moreover, the detected redshifts have a mean value of $z\sim1$ \citep{Bi2018}.

In recent years, finding such bursts with low luminosity and softer spectral peak energies has been difficult because of instruments designed to trigger on GRBs with high keV energies. However, some Swift GRBs, such as GRB 201015A, can be classified as XRFs \citep{Patel2023}.
With the advent of the X-ray mission \textit{Einstein Probe} (EP), launched in January 2024, the population of XRFs is growing. In particular, the Wide-field X-ray Telescope (WXT) on board EP is characterized by an extremely large field of view (3600 $deg^2$), good sensitivity, and a soft energy band (0.5-4 keV, \citealt{Yuan2022, Yuan2025}).

In this work, we present an extensive dataset of the afterglow of EP241021a, a fast X-ray transient discovered by EP. Since there were no detections by Fermi-GBM and Konus-Wind, this transient is consistent with XRF properties (see Section \ref{subsec:prompt}). In Section \ref{subsec:afterglow_dataset} we introduce the X-ray, optical, and radio datasets. Their spectral and temporal behavior is studied phenomenologically in Section \ref{sec:phenom}. In Section \ref{subsec:phys_modelling} we physically interpret the complete light curve and spectra, placing the features we obtained from the physical modeling in the context of collapsars in Section \ref{sec:discussion}. Finally, the conclusions are given in Section \ref{sec:conclusions}.

\section{Observations}
\label{sec:dataset}

\subsection{Prompt emission}
\label{subsec:prompt}

EP241021a was detected by the EP-WXT on October 21, 2024 \citep{EP_GCN}. The transient started at 2024-10-21T05:07:56 (UTC, T$_{0}$ hereafter) and had a duration of about 100 seconds. The average 0.5-4 keV spectrum is well fit by an absorbed power law with a photon index of $1.8$ and a column density fixed at the Galactic value of $5\times10^{20} \rm cm^{-2}$, with an average unabsorbed 0.5-4 keV flux of $3.31^{+1.26}_{-0.86} \times 10^{-10} \rm erg \ s^{-1} \ cm^{-2}$ \citep{Shu2025}. This translates into a fluence in the 2-30 keV band of $\sim6\times10^{-8} \rm erg \ cm^{-2}$. 

There is no high-energy (gamma-ray) counterpart of this event. In particular, Fermi-GBM operated as expected at that time, and the position of the transient was not Earth-occulted. We extracted the Fermi-GBM light curve of the nb detector, (the nearest to the source position) using the Fermi Gamma-ray Data Tools \citep{GDT-Fermi}. Modeling the background as a first-order power law, we find a rate of two counts per second. This leads to a 3$\sigma$ upper limit \citep{Gehrels1986} on the fluence in the 30-400 keV band assuming a duration of 100 s of: $1\times10^{-6} \rm erg \ cm^{-2}$ for a long GRB spectrum (Band function with $\alpha=-1$, $\beta=-2.5$, and peak energy $E_p$=300 keV); $5\times10^{-7} \rm erg \ cm^{-2}$ for an XRR spectrum (Band function with $\alpha=-1$, $\beta=-2.5$, and peak energy $E_p$=70 keV); and $2\times10^{-7} \rm erg \ cm^{-2}$ for an XRF spectrum (Band function with $\alpha=-1$, $\beta=-2.5$, and peak energy $E_p$=30 keV). In addition, Konus-Wind also observed the whole sky at the time of the transient but did not detect the source \citep{KW_GCN}.

Following \cite{DAlessio_2006} and depending on the ratio of the fluences in the bands $\rm F(2-30 keV)/F(30-400 keV)$, we classify this transient as either a GRB ($\rm F(2-30 keV)/F(30-400 keV)<0.3$), an X-ray-rich GRB ($\rm 1<F(2-30 keV)/F(30-400 keV)<0.3$), or an XRF ($\rm F(2-30 keV)/F(30-400 keV)>1$). Assuming an XRF spectrum, the upper limit on the 30-400 keV and the 2-30 keV fluences places this transient in the XRR and XRF population, with $\rm F(2-30 keV)/F(30-400 keV) \geq 0.3$.

The spectroscopic redshift is observed to be 0.75 \citep{GCN_GTC_z, GCN_VLT_z, GCN_Keck_z}. This gives a 0.5-10 keV luminosity of $1.3 \times 10^{48} \rm erg \ s^{-1}$, which is consistent with low-luminosity GRBs, and very similar to EP240414a luminosity in the same band \citep{Sun_2024_EP240414a}. 

Due to the uncertainty on the photon index $\Gamma=1.8^{+0.57}_{-0.54}$ found in \citet{Shu2025}, it is difficult to constrain the peak energy $E_{peak}$ of the spectrum. We find that for $E_{peak} \geq 30$ keV ($E_{iso}\sim10^{52}$ erg), the transient is marginally consistent with the Amati relation \citep{Willingale2017}.

\subsection{Afterglow}
\label{subsec:afterglow_dataset}

\subsubsection{Einstein Probe FXT}
\label{subsubsec:EP_data}

The Follow-up X-ray Telescope \citep[FXT,][]{Chen2020SPIE} on board EP operates in the 0.3--10 keV band and consists of two co-aligned modules (FXT-A and FXT-B). EP241021a was observed for fifteen total epochs by EP-FXT, starting from 2024-10-22 to 2025-01-08. Data were reduced using the recommended procedures from the FXT Data Analysis Software v1.10: particle event identification, pulse invariant conversion, grade calculation and selection (grade $\lesssim$ 12), bad- and hot-pixel flagging and selection of good time intervals using the housekeeping file to produce cleaned event files. The cleaned event files were then filtered for background flares, after which images and spectra were extracted for further analysis.

For source detection, we performed the recommended checks for EP-FXT data\footnote{http://epfxt.ihep.ac.cn/downloads/FXT\_Users\_Guide\_v1\_00-2.pdf}, employing an S/N (signal-to-noise ratio) threshold of three. The source was detected at nine epochs out of the fifteen. We estimated the source count rates from a 1 arcmin circle (corresponding to 95\% of the FXT point spread function - PSF), centered on the e-MERLIN position (see Section \ref{subsec:merlin}), and the background rates from an annulus with an inner radius of 1.2 and outer radius of 2.5 arcmin (thus, away from the source PSF and at least three times larger in area). Upper limits for the six non-detections were derived by taking the single-sided 3$\sigma$ Poissonian confidence limit following the procedure from \cite{Gehrels1986}.

We calculated an unabsorbed counts-to-flux conversion factor by assuming the best-fit spectral parameters of the observation at 8 days post-event (06800000186), where we had the most significant late-time detection of the X-ray counterpart. The spectra for this epoch were fit with an absorbed power law, with a single absorption component fixed to the Galactic neutral column in the direction of EP241021a ($5 \times 10^{20} \rm cm^{-2}$) and from which we derived a best-fit photon index $\Gamma$ of $1.81\pm0.30$. This yields a counts-to-flux conversion factor of $3.1\times10^{-11}$.
We also individually fit all the FXT observations with a detection and found a photon index consistent within errors with $1.81\pm0.30$, indicating negligible spectral evolution.
When we had a detection in both modules, we derived the final count rate as the average of those from the FXT-A and FXT-B telescopes. The results of the FXT data reduction are compiled in Table \ref{table:xray}. We jointly fit the spectra with the highest number of counts up to 6.0 days, with an absorbed power law (the single absorption component fixed to the Galactic neutral column), deriving a best-fit photon index of $\Gamma = 1.92\pm0.22$. Finally, we also jointly fit all the FXT spectra from 1.5 to 8.6 days with the same model, deriving a best-fit photon index of $\Gamma = 1.80\pm0.20$, in agreement with the analysis of \cite{Shu2025}.

\subsubsection{LBT}

We observed the optical counterpart of EP241021a with the Large Binocular Cameras \citep[][]{Giallongo2008a} mounted on the Large Binocular Telescope (LBT; Mt. Graham, AZ, USA) and obtained simultaneous $r^\prime$ and $ z^\prime$ imaging on 2024-10-28, at midtime UT 06:25:00, 7.05 days after the burst (Program IT-2024B-023, PI Elisabetta Maiorano). Observations were performed under an average seeing of 0\farcs85 for a total of 600 s in each filter. Imaging data were reduced using the dedicated data reduction pipeline \citep{Fontana2014a}. All data were analyzed by performing aperture photometry using DAOPHOT and APPHOT under PyRAF/IRAF, and calibrated against the SDSS DR12 catalog magnitudes of brighter nearby stars \citep{Alam2015a}. The optical afterglow was well detected in all bands. We measure an AB magnitude of $r^\prime = 21.78 \pm 0.02 $ and $z^\prime = 21.25 \pm 0.03$, not corrected for the foreground Galactic extinction $A_{\lambda}$ ($A_{\lambda}$=0.11 in the $r^\prime$ band and $A_{\lambda}$=0.065 in the $z^\prime$ band).

\subsubsection{GTC}
Near-IR observations with the 10.4-meter Gran Telescopio Canarias (GTC) in La Palma (Canary Islands, Spain) in the JHK-bands were conducted with the IR multi-object spectrograph EMIR (Espectrografo Multiobjeto Infra-Rojo) on two epochs (November 13, 2024, and February 3, 2025) with total on-source times of 350s (J), 462s (H), and 588s (Ks) each time, under program GTCMULTIPLE7B-24B. Data were reduced using the standard EMIR pipeline and calibrated using the 2MASS catalog. Observations are reported in Table \ref{table:optical}, not corrected for $A_{\lambda}$, with $A_{\lambda}=0.037, 0.02$ and 0.02 in the J, H, Ks bands, respectively.

\subsubsection{CAHA}
Optical observations with the 2.2-meter Calar Alto Telescope (CAHA) in southern Spain were conducted with the Calar Alto Faint Object Spectrograph (CAFOS) in the g$^\prime$r$^\prime$i$^\prime$z$^\prime$- Sloan bands on December 7 and 8, 2024, under program 24B-2.2-012, with on-source exposures of 840s g', 630s r', and 360s i$^\prime$ (December 7) and 900s r$^\prime$ and 900s i$^\prime$ (December 8). Data were reduced using the standard IRAF routine and calibrated using the Sloan Digital Survey Photometric Catalog. Observations are reported in Table \ref{table:optical}, not corrected for $A_{\lambda}$, with $A_{\lambda}=0.17, 0.11$ and 0.09 in the g$^\prime$, r$^\prime$, and i$^\prime$ bands, respectively.

\subsubsection{e-MERLIN}
\label{subsec:merlin}
We began observations of EP241021a on October 24, 2024, under our approved Target-of-Opportunity e-MERLIN (enhanced Multi Element Remotely Linked Interferometer Network) project CY18212 (PI: Gianfagna) and monitored the source for four epochs (see Table \ref{table:radio}) at 5.5 GHz. Each run lasted $\sim$12 hours, alternating between target and phase calibrator. Data were calibrated with the e-Merlin pipeline \citep{2021asclsoft09006M}, and then imaged in {\tt{CASA}} (Common Astronomy Software Applications, \citealt{TheCASATeam_2022}), reaching a typical angular resolution of $\sim$150$\times$50 milli-arcsec, and an RMS of $\sim$50 $\mu$Jy/beam. Observation results are reported in Table \ref{table:radio}.

\subsubsection{ATCA}
We monitored the source for a total of six epochs with the Australia Telescope Compact Array (ATCA) starting on October 29, 2024, through our approved program C3638 (PI: Gianfagna) and the shared program CX585 (PIs: Carotenuto, Gianfagna, Troja, Yao). Observations were carried out with the 16cm (2.1 GHz), 4cm (5.5 GHz and 9 GHz), and 15mm (16.7 GHz and 21.2 GHz) receivers, in phase referencing mode. Data were processed in {\tt{CASA}} \citep{TheCASATeam_2022}, with standard recipes. The combination of different array configurations during the approximately three-month campaign, and the equatorial declination of the target, made imaging challenging for some epochs. In such cases, resulting in an extremely elongated beam, we measured the source flux density through a direct fitting of the interferometer visibilities with the {\tt{UVMODELFIT}} task in {\tt{CASA}}. Otherwise, when imaging was feasible, we extracted the peak flux density in a region centered on the source position. Results are reported in Table \ref{table:radio}.

\subsubsection{GMRT}
We obtained director discretionary time (DDT) through the project ddtC407 (PI: Gianfagna) with GMRT (Giant Metrewave Radio Telescope). We performed two epochs with a total of 8 hours, divided among Band 3 (250-500 MHz), Band 4 (550-580 MHz), and Band 5 (1050-1450 MHz). Observations were conducted on December 30-31, 2024, and March 23, 2025. Data were processed with the {\tt{SPAM}} pipeline\footnote{https://www.intema.nl/doku.php?id=huibintema:spam:start}. For the first epoch, we reached an RMS of 155 $\mu$Jy/beam in Band-3, and 24 $\mu$Jy/beam in Band-5, while Band-4 failed to produce a usable image due to data corruption by RFI (radio frequency interference). We observed only in Band-5 for the second epoch, reaching an RMS of 38 $\mu$Jy/beam. No detection was achieved at either epoch (see Table \ref{table:radio} for the resulting upper limits).

\subsubsection{ALMA}
Another DDT was granted by the Atacama Large Millimeter/submillimeter Array (ALMA) with code 2024.A.00019.T (PI: Gianfagna). Observations were carried out in the period of December 28-30, 2024, for a total of 11 min on source in Band 1, with central frequency 40 GHz, and 2.5 hrs on source in Band 6, with central frequency 233 GHz. 
The data have been calibrated using the pipeline (version 2024.1.0.8) and images have been obtained using the task {\it{tclean}} in {\tt{CASA}} (version 6.6.1.7; \citealt{TheCASATeam_2022}). 
The multifrequency synthesis of the data over the full bandwidth 
($\sim$ 7.5 GHz) resulted in a rms of 15 $\mu$Jy/beam (4.46 $\times$ 3.525 arcsec) in Band 1, and of 10 $\mu$Jy/beam (0.7 $\times$ 0.6 arcsec) in Band 6. 
We obtained in both bands detections of point-like sources with a S/N larger than 10, and the corresponding flux densities (with their error) are reported in Table \ref{table:radio}.
As a quality check, we also obtained images of the four different 
spectral windows, and the source does not show any significant in-band 
variation in flux.  

\section{Analysis}

\begin{figure}
   \centering
   \includegraphics[width=\hsize]{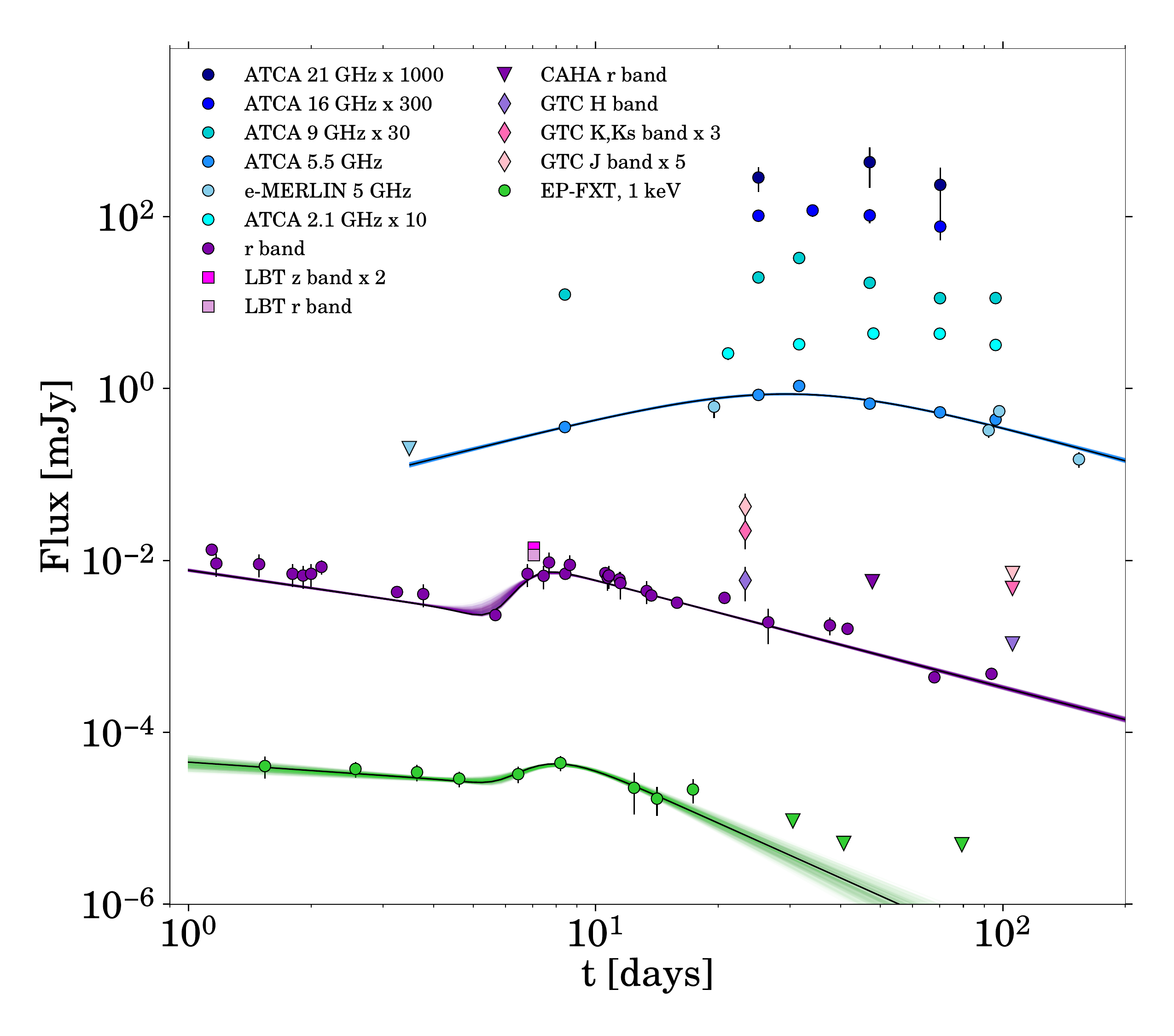}
      \caption{EP241021a broadband light curve. FXT observations are represented with green data points; optical observations are represented in shades of purple and pink; and radio data are represented in shades of blue. Inverted triangles indicate upper limits. The optical data represented with circles are taken from the literature \citep{Busmann2025, Aryan2025, Fu_GCN,  Fu2_GCN, Li_GCN, Jin_GCN, Pan_GCN, Ror_GCN, Kumar_GCN, Busmann_GCN, Bochenek_GCN, Bochenek2_GCN, Bochenek3_GCN, Freeburn_GCN, Freeburn2_GCN, QuirolaV_GCN,  Moskvitin_GCN, Moskvitin2_GCN, Moskvitin3_GCN, Schneider_GCN}. The solid black lines and shaded-colored regions represent the best fit and the 500 best likelihood fits of the data using power law models.}
         \label{Fig:lightcurve}
   \end{figure}

The broadband light curve of the afterglow of EP241021a is presented in Fig. \ref{Fig:lightcurve}. While the optical and X-rays are observed to decrease with time, the radio band shows a rising phase and a peak around 30 days.

\begin{figure*}
   \centering
   \includegraphics[width=0.45\linewidth]{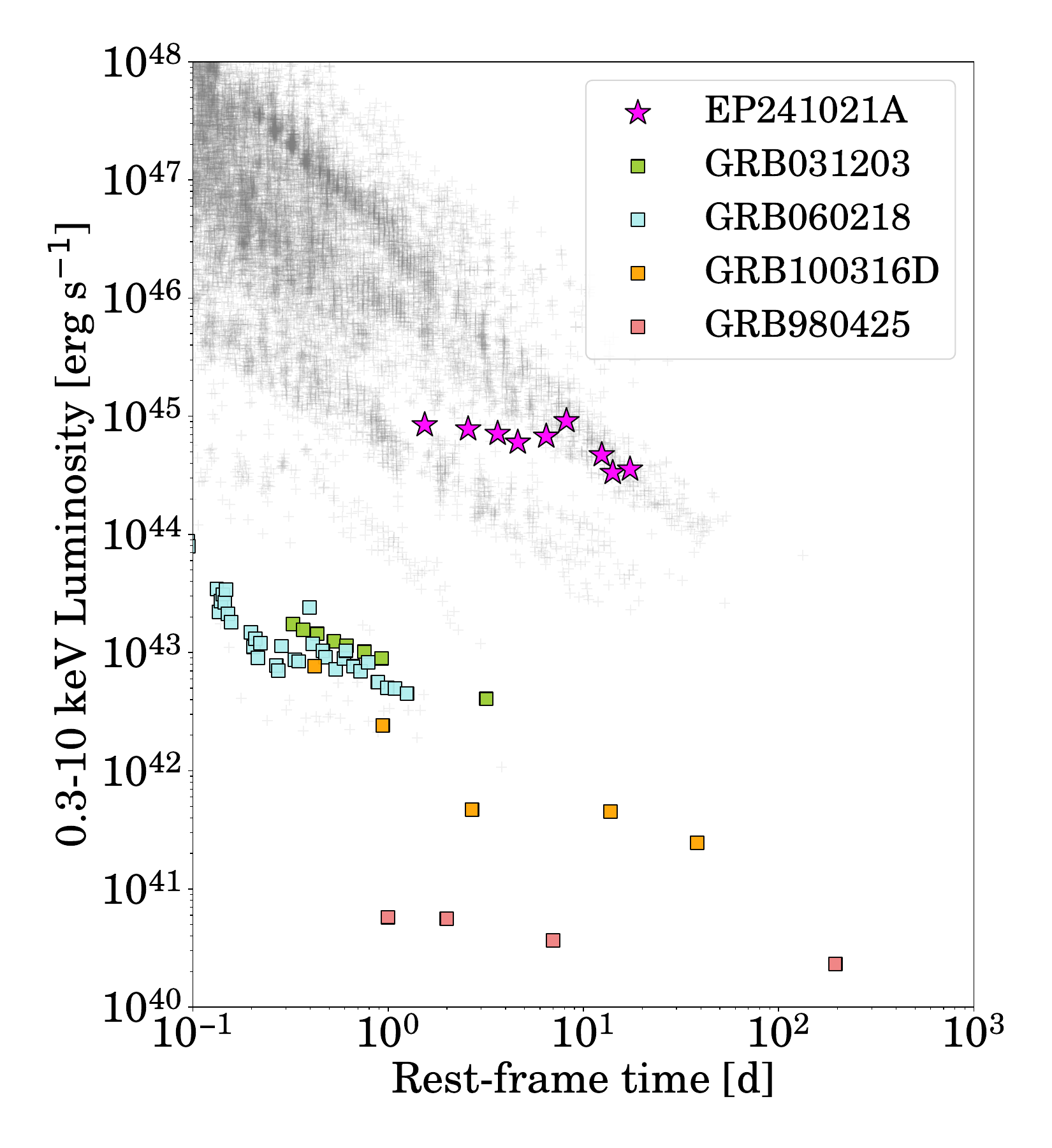}
   \includegraphics[width=0.45\linewidth]
   {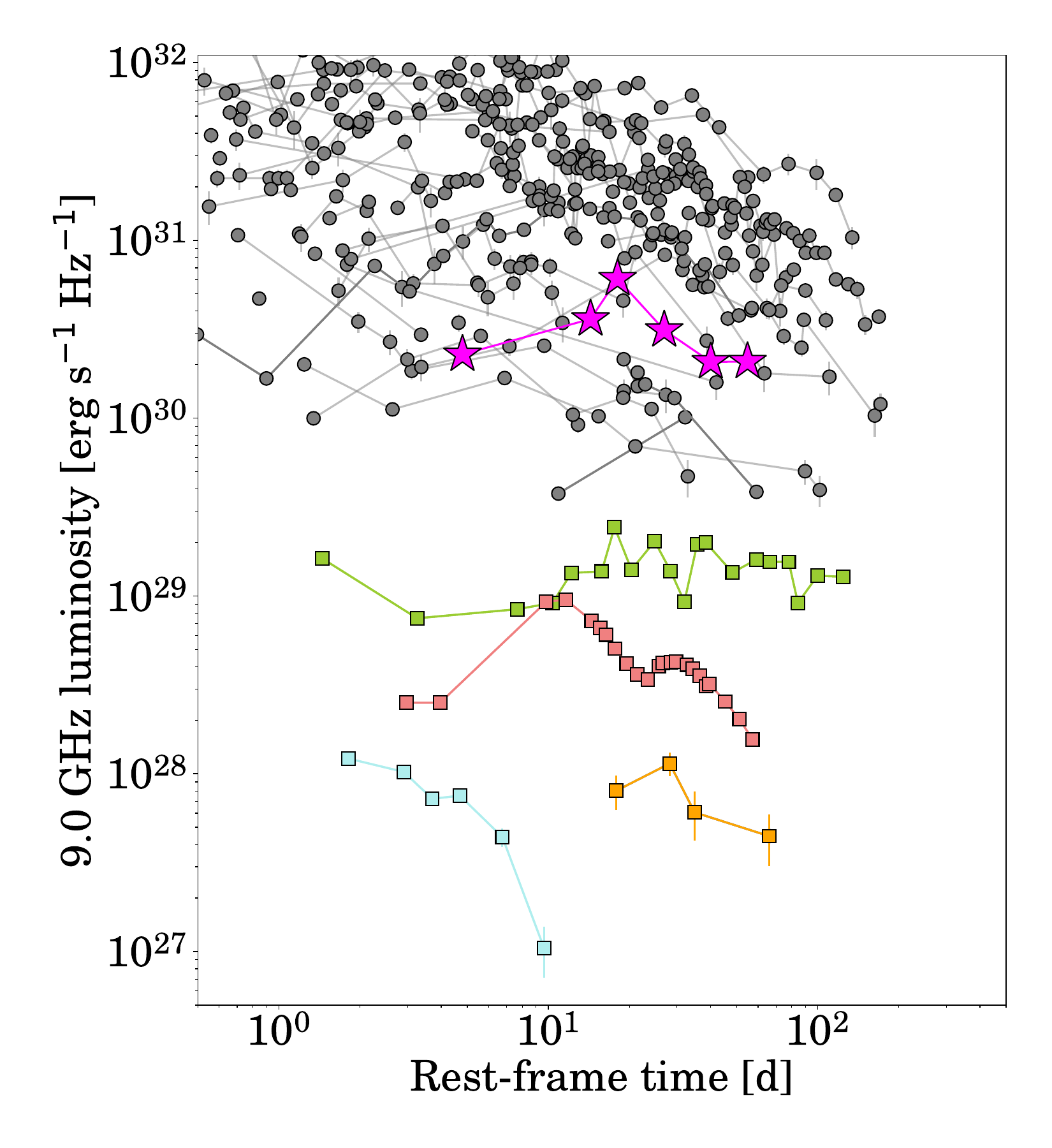}
      \caption{Left panel: X-ray luminosities in the 0.3-10 keV band of Swift-detected GRBs \citep{Evans2007, Evans2009}. Right panel: Radio afterglow luminosities at 9 GHz of the GRB catalog presented in \cite{ChandraFrail2012}. EP241021a is represented in magenta stars, while low-luminosity GRBs, such as GRB 060218 \citep{Campana2006, Soderberg2006}, GRB 100316D \citep{Starling2011, Margutti2013}, GRB 980425 \citep{Pian1999}, and GRB 031203 \citep{Watson2004} are represented with light blue, yellow, light red, and green squares, respectively, in both panels.
      }
         \label{Fig:radio_catalogue}
   \end{figure*}

The X-ray afterglow luminosity is observed to span the order of $10^{44}-10^{45} \rm erg \ s^{-1}$. At early times before the bump, the emission was placed at the lower end of cosmological GRBs, between the majority of the GRB population and low-luminosity GRBs, such as GRB060218, GRB100316D, and GRB980425 (see Fig.\ref{Fig:radio_catalogue}, left panel). 
The radio luminosity further supports this association (see Fig.\ref{Fig:radio_catalogue}, right panel). Instead, the late X-ray luminosity is consistent with the standard GRB population. This could suggest that the EP241021a X-ray emission after the peak is likely due to a relativistic jet, while the early X-ray and the full radio light curve could be due to a less energetic component. As an example, nonrelativistic flows (such as SNe type Ic) can be clearly rejected as they have luminosities on the order of $10^{28} \rm erg \ s^{-1} Hz^{-1}$ at 9 GHz.

In the following sections, we analyze the EP241021a light curve and broadband spectrum using a phenomenological approach and a physical interpretation. The fits are performed using the {\tt{Python}} package {\tt{Dynesty}} \citep{Higson2019, Speagle2020, koposov2024}, and the corner plot is made using {\tt{corner}} \citep{corner}. The results are always presented as median, 16th, and 84th percentiles.

\subsection{Phenomenological analysis}
\label{sec:phenom}

\subsubsection{Light curve}
\label{subsec:phenom_lc}

We assumed a time and spectral dependence for the flux of the type $F\propto t^{\alpha} \nu^{\beta}$ and fit the light curve in X-rays, the r band, and at 5-5.5 GHz with smoothly broken power laws. 
In the case of the radio band, we find slopes $\alpha_{r,1} = 1.2\pm0.1$ and $\alpha_{r,2}=-1.2\pm0.1$ and break time $t_{r,b} = 28.4^{+2.4}_{-3.7}$d.
In the case of the optical, we find that the early ($t<6$ d) light curve has a slope of $\alpha_{o,0} = -0.70 \pm 0.03$, then there is a fast rising slope of $\alpha_{o,1} = 4.6 \pm 1.5$, a bump with $t_b=8.1^{+0.9}_{-0.7}$d and a decreasing slope of $\alpha_{o,2} = -1.26\pm0.05$ (all in agreement with \citealt{Busmann2025}). 
In the X-rays, the early slope is $\alpha_{x,0} = -0.3^{+0.2}_{-0.1}$, then $\alpha_{x,1} = 3.1 \pm 1.2$, a bump with $t_b=7.9^{+0.7}_{-0.6}$d and a decreasing slope of $\alpha_{o,2} = -2.3\pm0.5$. The slopes $\alpha_1$ and peak times of the optical and X-ray light curves agree within 1$\sigma$, while the slopes $\alpha_2$ seem to suggest a milder decrease in the optical flux. The X-ray slope is consistent with a post-jet break phase. 

Focusing on optical and X-rays, we first considered the early afterglow emission, up to about 7 days. The different slopes suggest that, assuming a synchrotron emission, a break frequency is between the two bands. 
According to the afterglow theory, in an ISM (interstellar medium) environment \citep{Sari1998}, the flux at high frequencies is expected to decrease faster with time than at lower frequencies, contrary to what is observed. However, in the case of a wind environment \citep{Chevalier2000}, the optical light curve can be steeper than the X-ray, with a difference of 1/4.
EP241021a does show a steeper slope for the optical band, suggesting a wind environment, but the difference from the X-rays is $\sim0.4$. Such a large difference can be explained by continuous energy injection due to refreshing material. The latter can produce milder decays. We thus use this model to explain the early afterglow emission in Section \ref{subsec:early-time-ag}. 
The bump and the ensuing decay observed in optical and X-rays will be discussed in more detail in Section \ref{subsec:rebrightening}.

\subsubsection{Spectrum}
\label{subsec:phenom_spectrum}

\begin{figure}
   \centering
   \includegraphics[width=\hsize]{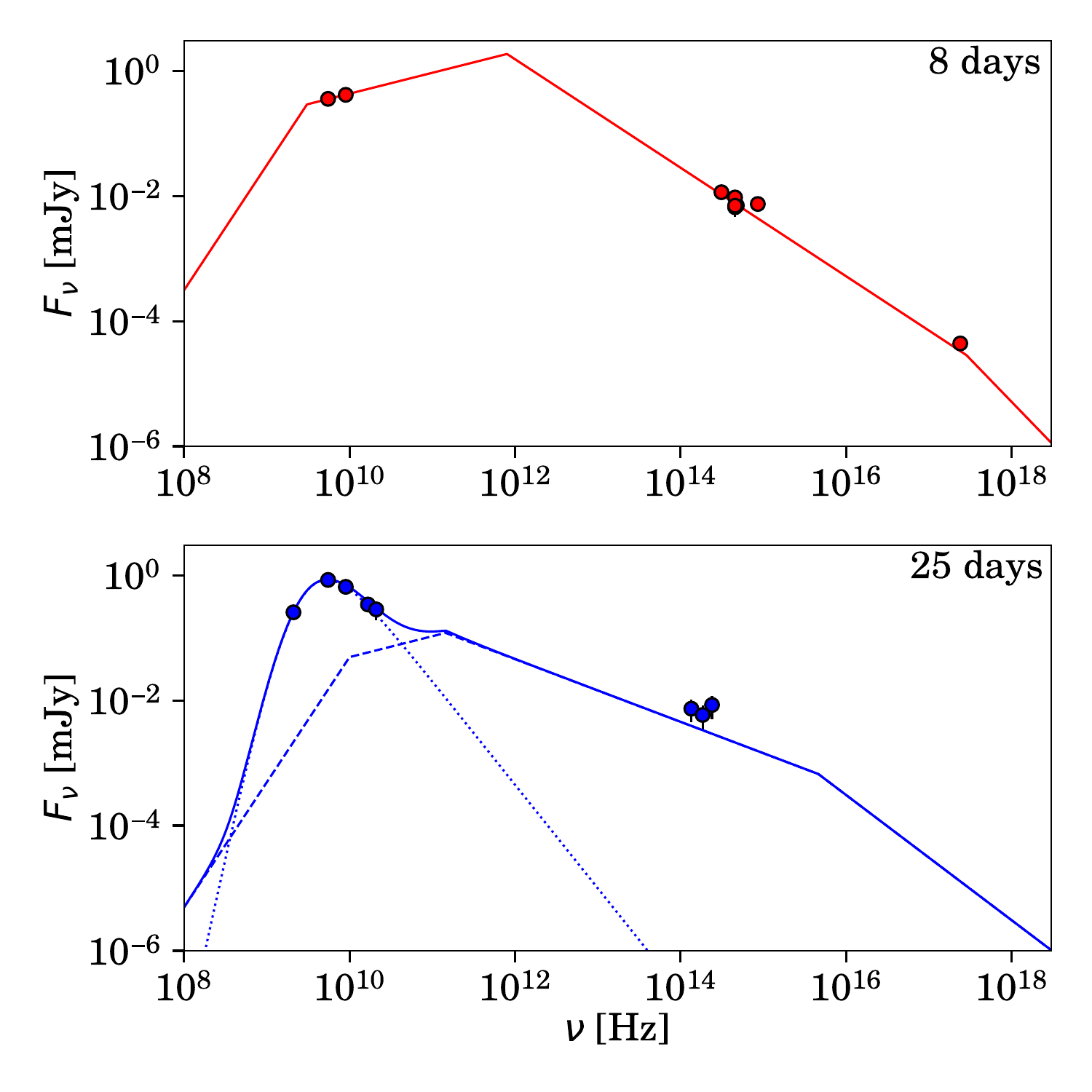}
      \caption{EP241021a broadband spectrum at $\sim$8 d (top panel) and $\sim$25 d (bottom panel). In the top panel, the solid lines represent a fit of two broken power law models. In the bottom panel, the solid line represents the sum of the radio and optical components, represented with dotted and dashed lines, respectively.}
         \label{Fig:broadband_spectra}
   \end{figure}

We repeated the same analysis for the broadband spectrum. We used a three-broken power law model, fixing the slopes of each power law to the slopes expected in the synchrotron modeling. In particular, from low to high energies, the slopes are $2$, $1/3$, $-(p-1)/2$, and $p/2$, in the assumption of slow cooling. The fitted parameters are then $p$ and three break frequencies: the synchrotron self-absorption $\nu_{sa}$, the injection $\nu_m$, and the cooling $\nu_c$ frequencies. The spectra at $\sim$8 d and $\sim$25 d are represented in Fig.\ref{Fig:broadband_spectra}. At 8 days, we could only get an upper limit on the synchrotron self-absorption frequency, $\nu_{sa}<5.5$ GHz. The same occurred for $\nu_c$, for which we obtained a lower limit of $\sim 2.5 \times 10^{17}$ Hz.
We find $p=2.75^{+0.08}_{-0.10}$, $\log(\nu_m) = 11.93^{+0.11}_{-0.10}$, and a normalization $A=1.89^{+0.14}_{-0.17}$ mJy. 
The value of p is in agreement with the one estimated from the X-ray spectrum (Section \ref{subsubsec:EP_data}): assuming that X-rays are below $\nu_c$, this leads to a spectral slope $\beta = \Gamma - 1 = (p-1)/2 = 0.8 \pm 0.2$, and $p=2.6\pm0.4$. 
At 25 days, the spectral peak appeared to be at $\sim 5$ GHz. Assuming that this is still $\nu_m$, this would correspond to a time dependence of $\sim t^{-4}$, hardly explainable with synchrotron theory. We note that all the radio spectra, except the one at 8 days, exhibit the same bell shape. This is also reported in \citet{Shu2025}. Hence, radio emission after $\sim$25 days is likely produced by a component separated from the one responsible for the optical and X-ray afterglow.
For this reason, we fit the 25 d broadband spectrum with the sum of two different components, one for the radio and one for the optical. We discuss the radio spectrum in detail later, while here we focus on the optical. We always assumed a three-broken power-law model, imposing that $\nu_m$ evolves from 8 days onward following a time dependence of $\nu_m \propto t^{-3/2}$, which is valid both for wind and ISM environments. The fit is represented with the dashed line in the second panel of Fig. \ref{Fig:broadband_spectra}. We find $p=2.05^{+0.10}_{-0.04}$, $\log(\nu_m) = 11.19^{+0.11}_{-0.10}$, and a normalization $A=0.12^{+0.02}_{-0.03}$ while $\nu_{sa}$ and $\nu_c$ are unconstrained.
This result clearly shows the presence of a separate component in the radio band (see Fig.\ref{Fig:broadband_spectra}, bottom panel and the following text), a result independent of the value of $\nu_{sa}$.

\begin{figure}
   \centering
   \includegraphics[width=\hsize]{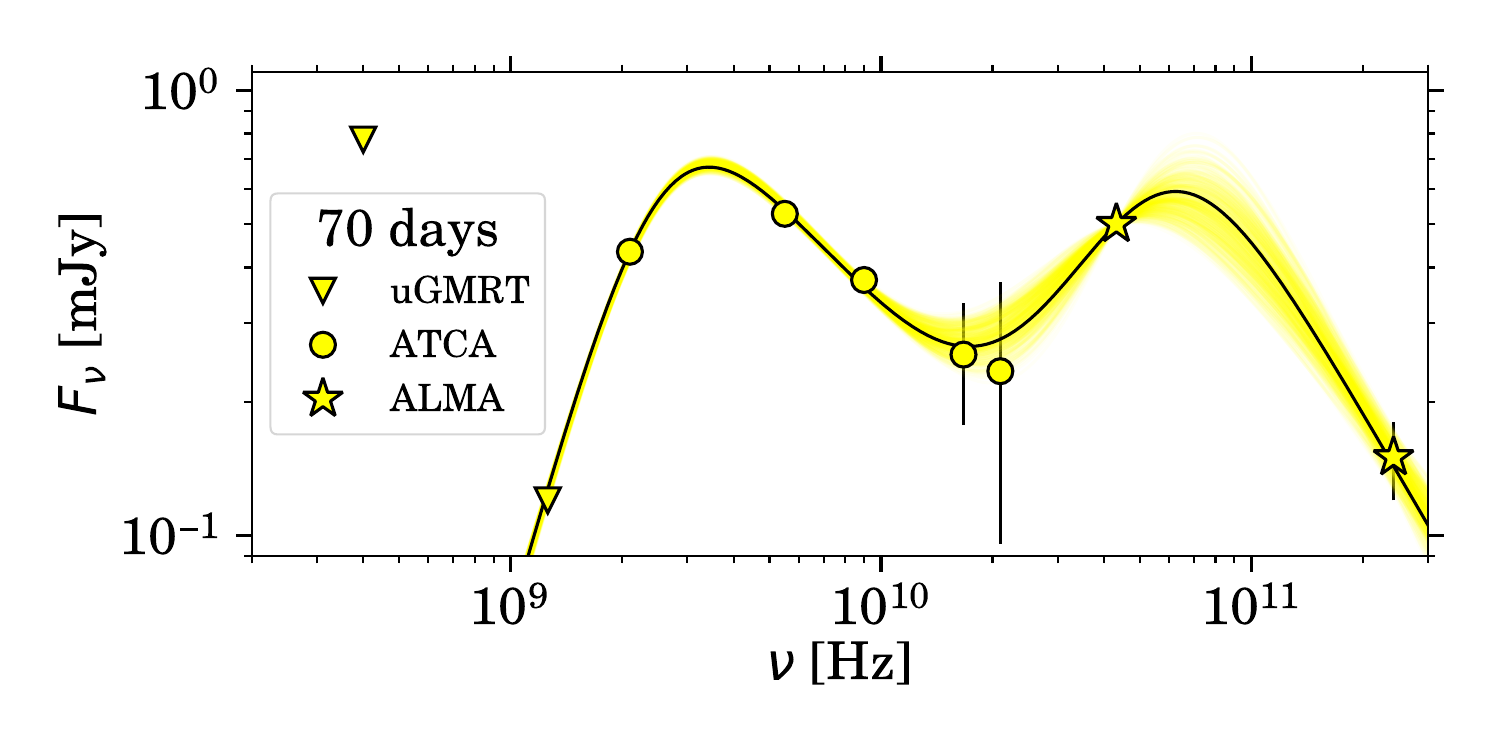}
      \caption{EP241021a radio spectrum at 70 days. The 5 sigma upper limits from uGMRT are represented as triangles; ATCA observations are represented with dots; and ALMA observations are represented with stars. The spectra are fit with a broken power-law model, the colored regions represent the 500 highest-likelihood fits, while the black line represents the best fit. }
         \label{Fig:radio_spectrum_PL}
   \end{figure}

We now turn to the radio spectrum, which we fit with one smoothly broken power law. The fitted parameters are the two slopes $\beta_1$ and $\beta_2$ before and after the break frequency $\nu_b$, and the peak flux $F_b$. We show the epochs at 25 d in Fig.\ref{Fig:broadband_spectra}, bottom panel, and 70 d in Fig.\ref{Fig:radio_spectrum_PL}. The results are reported in Table \ref{table:fit_res_phenom}. Interestingly, at 70 days, the spectrum is double-peaked, with a second peak emerging in the ALMA bandpass, and we use the sum of two smoothly broken power laws to fit the data. In the table, we report the results of the fit for both components, represented in Fig.\ref{Fig:radio_spectrum_PL}. 

\begin{table*}[h!]
\caption{Phenomenological fit of the radio spectra.} 
\label{table:fit_res_phenom}     
\centering                       
\begin{tabular}{c c c c c c}      
\hline\hline               
  & 25 d & 30 d & 48 d & 70 d & 70 d II\\   
\hline                       
   $F_b$ [mJy]   & 1.6$\pm0.1$ & 2.3$\pm0.1$ & 1.3$\pm0.1$ & 1.1$\pm0.1$ & $0.94^{+0.05}_{-0.08}$ \\      
   $\log$($\nu_{b}$ [Hz])  & 9.7$\pm0.1$ & 9.74$^{+0.05}_{-0.04}$ & 9.6$\pm0.1$ & 9.42$\pm0.03$ & $10.60_{-0.04}^{+0.03}$ \\
   $\beta_1$ & 2.4$\pm0.4$ & $2.0\pm0.2$ & 1.5$^{+0.4}_{-0.7}$ & 2.8$\pm0.2$ & $2.78_{+0.31}^{-0.20}$ \\
   $\beta_2$ & -1.2$\pm0.2$ & -1.4$^{+0.1}_{-0.2}$ & -0.8$^{+0.3}_{-0.2}$ & -0.9$^{+0.1}_{-0.2}$ & $-1.2_{-0.1}^{+0.2}$ \\
\hline      
\end{tabular}
\tablefoot{The first column lists the parameters of the fit, the following columns list the mean time at which each spectrum was taken. At 70 d, we find a double-peaked spectrum, and the two components are reported in the last two columns.}
\end{table*}

The power law spectral shape mentioned above confirms a synchrotron origin also for the EP241021a radio afterglow. A rising slope between  $\sim$2-2.5 is expected when synchrotron self-absorption is dominant; thus, the break frequency $\nu_b$ corresponds to $\nu_{sa}$. In particular, a slope of $\beta_1\approx 2.5$ is expected  when the electron injection frequency is below the self-absorption frequency ($\nu_m < \nu_{sa}$). This order is further supported by the absence of a $\nu^{1/3}$ regime above the break frequency. In fact, in our observations, higher frequencies are following a power law with a slope between -1.4 and -0.8, and are all in agreement within 2$\sigma$.
In the case of slow cooling synchrotron emission, such slopes would correspond to $-(p-1) / 2$, leading to a mean value of $p$ of $3.1\pm0.5$.

The break frequency $\nu_{sa}$ has a value of 5 GHz at 25 d. If we assume a GRB synchrotron spectrum, in the case of a wind environment, $\nu_{sa}$ is given by
\begin{equation}
    \nu_{sa} = 1\times10^{11} \left(\frac{1+z}{2} \right)^{-2/5} \left(\frac{\epsilon_e}{0.1} \right)^{-1} \left(\frac{\epsilon_B}{0.1} \right)^{1/5} \left(\frac{E_0}{10^{52}} \right)^{-2/5} A_*^{6/5} t_d^{-3/5} \ \rm Hz.
\end{equation}
The main contributions to $\nu_{sa}$ are by $\epsilon_e$ and $A_*$ \citep{Chevalier2000}. Assuming $E_0=10^{52}$erg and $\epsilon_B=0.1$, having $\nu_{sa}=5$GHz would mean an extremely high density of $A_*=6\times10^3 \rm cm^{-1}$ for $\epsilon_e=0.1$, in contrast with the highest end of the GRB distribution of 10 \citep{Aksulu2022}. 
In the case of a GRB in an ISM, $\nu_{sa}$ is given by
\begin{equation}
     \nu_{sa} = 1\times10^{8} \left(1+z \right)^{-1} \left(\frac{E_0}{10^{52}} \right)^{1/5} n_0^{3/5} \left(\frac{\epsilon_e}{0.1} \right)^{-1} \left(\frac{\epsilon_B}{0.1} \right)^{1/5} \ \rm Hz.
\end{equation}
Assuming the same values as above for $E_0$ and $\epsilon_B$, to get $\nu_{sa} =5$GHz, for $\epsilon_e=0.1$ an extremely high density $n_0\simeq10^5 \rm cm^{-3}$ is needed also in this case, with respect to typical ISM densities on the order of unity, or lower. 
Such high densities both in the wind and ISM environments are not only in disagreement with GRB studies, but also influence other afterglow traits, as the jet break time $t_{jb}$. The latter, in the case of a wind environment \citep{Chevalier2000} and assuming $E_0=10^{52}$erg,  $\epsilon_B=\epsilon_e=0.1$, a jet opening angle of $\theta_c=3^{\rm o}$, would be less than 1 s, instead of days.
The same is valid in the case of an ISM. For the same values of $E_0$, $\epsilon_e$, $\epsilon_B$ and $\theta_c$, $n_0\simeq10^5 \rm cm^{-3}$ implies a jet break time of $\sim$100 s \citep{Sari1999}. We also note that in Section \ref{subsec:phenom_lc}, from the temporal slopes of the optical and X-ray light curves, we find no indication of a jet break up to the bump at $\sim6$ days. For these reasons, we exclude the case of a GRB in either an ISM or a wind environment as the source of the radio emission.

\begin{figure}
   \centering
   \includegraphics[width=\hsize]{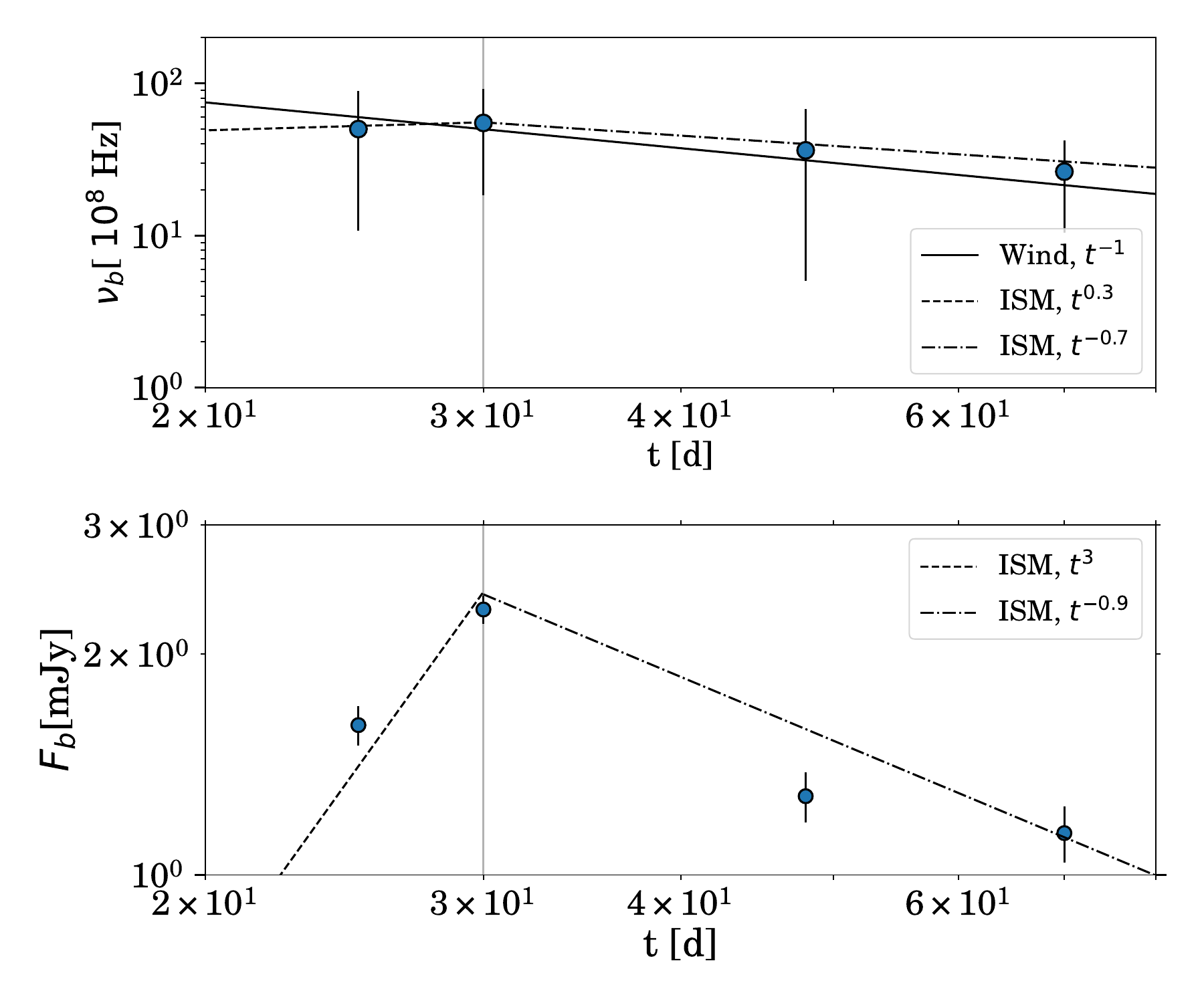}
      \caption{Top panel: Break frequency resulting from the fit as a function of time. The expected behavior is overplotted with a solid line in the case of a wind environment and with dashed and dot-dashed lines in the case of ISM. The vertical shaded line represents the assumed deceleration time. Bottom panel: Peak flux in the spectra as a function of time. The expected behavior as a function of time for the wind environment is constant, while that for the ISM is plotted in dashed lines.}
         \label{Fig:Flux_freq_peak_spec}
   \end{figure}

Having ruled out relativistic ejecta (either jetted or spherical) for the radio afterglow, we are left with mildly relativistic or nonrelativistic components (often modeled similarly, see \citealt{Piran2013}), in particular we focus on spherical ejecta (often called cocoons). A common example, which is often associated with GRBs, is SN radio emission. Such emission is characterized by a synchrotron self-absorbed spectrum, such as that which we observe. Being in a wind environment, the peak frequency of the spectrum ($\nu_{sa}$) is expected to decrease with time as $t^{-1}$, while the peak flux stays constant \citep{Chakraborti2011, Chevalier1998A}. In the case of an ISM, instead, the $\nu_{sa}$ of a spherical expansion is expected to increase as $t^{2/(p+4)}$, up to the deceleration time, and then decreases with time as $t^{-(3p-2)/(p+4)}$ \citep{Hotokezaka2015, Piran2013}. The flux rises across the whole spectrum before the deceleration time, and then decreases, with a slope that depends on $p$.
The break frequencies estimated from the fit are represented in the top panel of Fig.\ref{Fig:radio_spectrum_PL} as a function of time. The expected wind time dependence is represented with a solid black line, while the ISM model (assuming $p=2$) is represented with a dashed and a dot-dashed line, respectively, before and after the deceleration time. The latter is assumed to be 30 days (vertical shaded line), as this also corresponds to the peak of the radio emission.
The spectral peak fluxes are represented in the bottom panel of Fig. \ref{Fig:radio_spectrum_PL} (the expected behavior of the spectral peak flux in a wind is constant, and it is not plotted).
Focusing on the break frequencies, the ISM model seems to be preferred, even if the errors on $\nu_b$ are too wide to say that there is any time dependence at all. However, from Table \ref{table:fit_res_phenom} and bottom panel of Fig.\ref{Fig:Flux_freq_peak_spec}, it is clear that the peak flux $F_{p}$ increases up to 30 d and then decreases. This allows us to discard the spherical expansion in a wind, as the peak flux in the spectrum is expected to stay constant. 
Therefore, we identify a spherical expansion in a constant-density environment as the source of the radio emission. 

Even in the case of a cocoon scenario, we expect a large density. In fact, the synchrotron self-absorption in a cocoon in ISM \citep{Piran2013, Hotokezaka2015} is given by 
\begin{equation}
    \nu_{sa} = 1 \rm GHz (1+z)^{-1}\left(\frac{E}{10^{49}}\right)^{\frac{2}{3(p+4)}}n_0^{\frac{3p+14}{6(p+4)}}\left(\frac{\epsilon_B}{0.1}\right)^{\frac{2+p}{2(p+4)}}\left(\frac{\epsilon_e}{0.1}\right)^{\frac{2(p-1)}{p+4}}\beta_0^{\frac{15p-10}{3(p+4)}},
\end{equation}
at the deceleration time $t_{dec}$. Here, $E$ is the total kinetic energy of the ejecta at initial velocity $\beta_0$. Assuming that, at $t_{dec}$, $\nu_{sa}= 5 \rm GHz$, $\epsilon_e = \epsilon_B = 0.1$, $E=10^{49} \rm erg$, and $p=2$, for a $\beta_0=0.5$ the density has to be on the order of $10^2 \rm cm^{-3}$.

\subsection{Modeling all the EP241021a components}
\label{subsec:phys_modelling}

\subsubsection{Early-time afterglow}
\label{subsec:early-time-ag}

As mentioned above, the slopes of the optical and X-ray light curves suggest a refreshed shocks scenario. We used the model by \citet{Sari2000}, where the source is assumed to eject mass with a range of Lorentz factors $\gamma$, with $M(>\gamma) \propto \gamma^{-s}$. The fast ejecta produces a forward shock (FS) when it reaches the material surrounding the collapse at the deceleration radius. As it reaches the fast ejecta, the slow ejecta produces the refreshing effect and a reverse shock (RS), which propagates inward.
In a wind environment ($\rho \propto r^{-g}$ with $g=2$), the slopes of the forward and reverse shock light curves are a function of $p$ and $s$ \citep{Sari2000}. The peak flux of the spectrum, its frequency ($\nu_m$, in the assumption of slow cooling), and the cooling frequency $\nu_c$, depend on the initial energy and the maximum Lorentz factor of the ejecta ($E_0$ and $\gamma_0$), on the density of the environment through $A_*$, and on the microphysical parameters $\epsilon_e$ and $\epsilon_B$.

As priors, we chose uniform distributions for each parameter, in particular: $\gamma_0$ in [15, 100], $E_0$ in [$10^{48}$, $10^{51}$] ergs, $\epsilon_e$ and $\epsilon_B$ in [0.3, 0.001].
We fit the data up to $\sim6$d (before the bump), including the optical and X-ray light curves, and the 5 GHz upper limit at 3.8 days. We fix the values of $g=2$ (wind) and $p$. The latter was derived from the X-ray data in Section \ref{subsubsec:EP_data}, where we find a photon index of $\Gamma=1.92\pm0.22$ before 6 d. Assuming the X-rays are above the synchrotron cooling frequency $\nu_c$ at such early times, this leads to a spectral slope $\beta = \Gamma - 1 = 0.92 \pm 0.2$, and $p=1.85\pm0.39$. Since $p>2$, we fix $p=2$. 

The light curve fit is represented in Fig. \ref{Fig:lightcurve_fit_all}, dot-dashed line, and Fig. \ref{Fig:lightcurve_refreshed}. The radio, optical, and X-ray data are fit with the blue, purple, and green curves, respectively. The fit is done only on the data before the bump, shown in Fig.\ref{Fig:lightcurve_refreshed} by the vertical dashed line.  
In Fig.\ref{Fig:lightcurve_refreshed}, the solid line represents the sum of the contribution expected for the forward and reverse shocks, while the dotted and dashed lines represent the two components separately. The forward shock dominates the emission only at high energy.
The fit easily explains the early part of the afterglow, before the bump. The parameters we find are reported in Table \ref{table:fit_res}. 
The fastest ejecta with $\gamma_0 = 39$ ($\log(\gamma_0) = 1.59^{+0.27}_{-0.27}$) and $E_0 = 2.6\times10^{49}$ erg ($\log(E_0) = 49.42^{+0.51}_{-0.67}$), reach the external medium at a deceleration time of $\sim10$ sec. 
The observations in the optical and X-ray bands correspond to the time at which the ejecta with Lorentz factors between 7 and 9 reach the fast ejecta. Up to 6 days, during which this component dominates the emission, there is no evidence of a jet break; this implies a jet opening angle wider than 25 deg \citep{Chevalier2000}. Therefore, the beaming-corrected energy has a lower limit of $2.7\times10^{48}$ erg.

Once all of the refreshing material reached the rest of the ejecta, the refreshing phase ended, and the afterglow returned to a standard light curve with increased energy. The shaded lines in Fig.\ref{Fig:lightcurve_refreshed} after the vertical dashed line show the extrapolation of the model, assuming that the refreshing phase continues, with material with Lorentz factors of $\sim$2 reaching the deceleration radius at $\sim$5000 days. The fact that the refreshing ejecta are still relativistic at such late times is due to the decreasing density of the environment with distance, $n \propto r^{-2}$. 
Due to the bump, we are actually unable to identify the exact moment at which the energy injection stops. However, the X-ray upper limit at 89 d from \citet{Shu2025} suggests that the injection stops shortly after the bump, as early as 7 days, corresponding to Lorentz factors of $\sim7$. This model is represented with dot-dashed lines in Fig. \ref{Fig:lightcurve_refreshed}. 

Considering that there is no sign of a jet break before the bump, it is likely that we observe a wide ejecta (as also found in \citealt{Busmann2025, Yadav2025} with Lorentz factors in agreement with our $\gamma_0$), on axis. Moreover, a $\gamma_0\sim40$ places this component at the low end of the $\gamma_0$ distribution of long GRBs in a wind environment \citep{Ghirlanda2018}. Therefore, this ejecta is either a slow, baryon-loaded jet or a part of a structured jet. 
In the latter case, this wide component corresponds to the extended and less energetic wings (low-$\gamma_0$), which surround an energetic and uniform jet core. We adopt this explanation in the rest of the paper, as it also explains the bump after 6 days, as we describe in the following section.

\begin{table*}
\caption{Fit results for the broadband light curve. } 
\label{table:fit_res}     
\centering                         
\begin{tabular}{c c c c c}      
\hline\hline               
  & Refreshed shocks & Off-axis jet & Cocoon I & Cocoon II\\   
\hline                   
   $\log(E_{\rm tot} [\rm erg])$        & 48.43$^{a}$ & 51.17$^{+0.04}_{-0.03}$ & 51.52$^{+0.25}_{-0.29}$ & 50.93$^{+0.06}_{-0.09}$ \\
   \hline
   $\log(E_0 [\rm erg])$        & 49.42$^{+0.51}_{-0.67}$ & 54.50$^{+0.04}_{-0.03}$ &   &  \\
   $\log(\gamma_0)$   & 1.59$^{+0.27}_{-0.27}$ &  &  \\      
   $\log(\epsilon_e)$ & -0.58$^{+0.06}_{-0.08}$ & -0.60$^{+0.11}_{-0.07}$ & -1.35$^{+0.35}_{-0.33}$ & -0.51$^{+0.01}_{-0.01}$ \\
   $\log(\epsilon_B)$ & -1.30$^{+0.58}_{-0.83}$ & -1.36$^{+0.07}_{-0.06}$ & -0.89$^{+0.29}_{-0.25}$ & -0.53$^{+0.04}_{-0.02}$ \\
   $\log(A_* [\rm cm^{-1}])$        & -1.69$^{+0.62}_{-0.43}$ & -0.49$^{+0.04}_{-0.05}$ &  \\
   $s$                & 2.46$^{+0.51}_{-0.45}$ &  & \\
   $p$                & 2.0 & 2.016$^{+0.003}_{-0.005}$ & 2.02$^{+0.01}_{-0.02}$ & 2.27$^{+0.10}_{-0.15}$ \\
   $\theta_v$         &  & 0.108$^{+0.002}_{-0.004}$ & \\
   $u$          &  &  & 1.06$^{+0.02}_{-0.02}$ & 0.34$^{+0.01}_{-0.01}$ \\
   $\log(n_0 [\rm cm^{-3}])$        &  &  & 3.45$^{+0.25}_{-0.28}$ & 3.90$^{+0.10}_{-0.10}$ \\
\hline                                   
\end{tabular}
\tablefoot{
    In the first column, we list the parameters for each model: refreshed shock in the jet wings (second column), off-axis top-hat jet core (third column), and two cocoons describing the radio emission at $\nu<20$GHz (fourth column) and at $\nu> 20$GHz (fifth column), respectively. In the second row, $E_{\rm tot}$ indicates the beaming-corrected energy for the wings refreshed shock and off-axis jet core models, and the total kinetic energy for the two cocoon models.
    \tablefoottext{a}{Lower limit assuming a jet opening angle of 15 deg.}}
\end{table*}

\subsubsection{Re-brightening at 7 days}
\label{subsec:rebrightening}

There is a clear re-brightening at $\sim$7 d, present both in the optical and X-rays. Here we discuss different explanations.

A re-brightening in the light curve can be caused by density inhomogeneities (overdensity) in the medium surrounding the star. An increase in the ambient density should be followed by a decrease in radio luminosity, arising from the increase in synchrotron self-absorption, which does not seem to be present. We also note that, if the X-rays are above the cooling frequency $\nu_c$, they are not influenced by density \citep{Kumar2000, Freedman2001}, and the re-brightening is expected mainly in the optical. However, this is not the case here, as at 8 d we find $\nu_c>2.5\times10^{17}$ Hz (see Fig.\ref{Fig:broadband_spectra}).
The most compelling argument against density fluctuation is the fast rise of the bump, with a $\Delta t/t\approx 0.2$, much shorter than expected in such a case, with $\Delta t/t\approx 1$ \citep{Nakar2003, Nakar2007, VanEerten2009, Gat2013}.

The optical and X-ray bump is reminiscent of refreshed shocks, which we already introduced in Section \ref{subsec:early-time-ag}. Indeed, this is the preferred explanation for this bump in \citet{Busmann2025}. A fast re-brightening is expected if the slow ejecta, which catches up with the relativistic jet, is characterized by a single Lorentz factor \citep[GRB 970508, as an example]{Piro1998}. Another clear example of a light curve with refreshed shocks is GRB 030329, which shows multiple components reaching the relativistic jet at different times, before and after the jet break \citep{Granot2003}. The fact that the shocks reach the relativistic material before or after the jet break time influences the timescale of the resulting bumps in the light curve. In particular, one expects that before the jet break, the bump rising time $\Delta t$ should be on the order of $t$ \citep{Vlasis2011}, instead, after the jet break $\Delta t < t$. In the case of EP241021a, $\Delta t \sim 1$d, while $t\sim6$d, suggesting a post-jet break scenario. This, however, disagrees with the slope of the light curve before the bump, which does not show evidence of a jet break. For this reason, we do not prefer this explanation for the optical and X-ray bump. 

The absolute peak magnitude of the optical bump is -21.5, consistent with fast blue optical transients (FBOTs, \citealt{Drout2014}). An FBOT was found associated with another EP event, EP240414a \citep{Sun_2024_EP240414a, Vandalen_2024_EP240414a}. However, their spectra are typically blue, which is not the case of the EP241021a, see also \citet{Busmann2025}.

Finally, bumps in the optical light curves of GRBs are often identified as SNe. In this case, the rise time and the absolute magnitude are respectively too short and too bright to be explained as a SN, see \citet{Busmann2025} for an extended discussion.

After discarding all the scenarios above, we interpreted this bump as a signature of an off-axis jet. This modeling was already suggested for XRF 030723 \citep{Butler2005}, which has a fast bump in the optical light curve, hardly explained with a SN due to its short timescale \citep{Huang2004, Fynbo2004}.
We used {\tt{afterglowpy}} \citep{Ryan_2020, Ryan_2024} to model the light curve and, because of the short rise timescale, we assumed a top-hat jet, fixing the jet opening angle $\theta_c$ to 0.03 rad (1.7 deg), as also found in \citet{Huang2004}. Moreover, we assumed a wind environment, as indicated by the early data (see Section \ref{subsec:early-time-ag}).
The fitted parameter and their priors are the following: the viewing angle $\theta_v$ in [0, $\pi/2$]; the isotropic equivalent kinetic energy of the jet $E_0$ in [$10^{51}$, $10^{55}$] ergs; the reference number density at a radius of $10^{17}$ cm in [$10^{-1}$, $10^{3}$] cm$^{-3}$; the power law slope of the accelerated electrons $p$ in [2,3], the fraction of energy in the accelerated electrons $\epsilon_e$ in [0.3, 0.001]; and the fraction of energy in the magnetic field $\epsilon_B$ in [0.3, 0.001].

\begin{figure}
   \centering
   \includegraphics[width=\hsize]{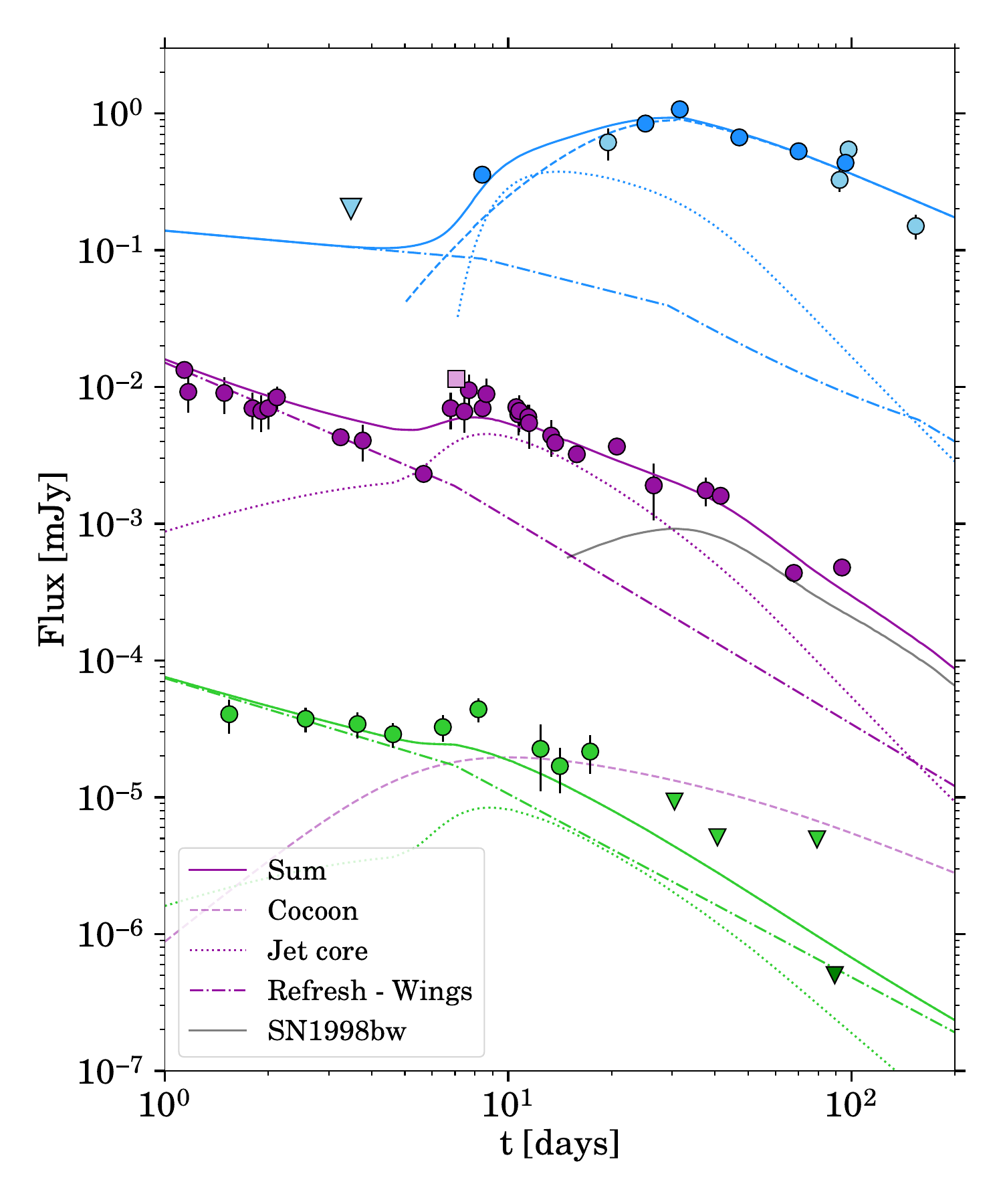}
      \caption{EP241021a broadband light curve modeling. The dataset is shown as in Fig.\ref{Fig:lightcurve}, with the addition of the dark green upper limit at $\sim$89 d from XMM Newton, reported in \citep{Shu2025}. The solid line represents the sum of an early refreshing phase (the wings of the jet, dot dashed line), the jet's off-axis core (dotted line), and a cocoon (dashed line). In the optical band, the expected emission of SN1998bw (shaded black line) is also included in the sum.
      The cocoon emission is mainly relevant for the radio band. The optical light curve is represented in the plot, while the X-ray light curve has a peak flux of $\sim 2\times10^{-8}$ mJy.}
         \label{Fig:lightcurve_fit_all}
   \end{figure}

The broadband fit is represented in Fig.\ref{Fig:lightcurve_fit_all}. We assume that the refreshing phase in the jet wings described in Section \ref{subsec:early-time-ag} ends at 7 d, this fit is represented with dot-dashed lines. The dotted line represents the off-axis jet, which mainly contributes to the bump.
The radio light curve is represented in blue in Fig.\ref{Fig:lightcurve_fit_all}. As shown in our phenomenological analysis (Section \ref{sec:phenom}), except for the early ($t<10$ d) data, the main contribution is due to another radio component, the cocoon, which we introduce in Section \ref{subsec:cocoon}.

Focusing on the early optical, X-ray, and radio data ($t<20$ d), the overall light curve behavior is well described by the sum of the early refreshed shock phase, taking place in a wide ejecta (the jet wings) surrounding the jet core, and the Top Hat core itself, seen off-axis. For the off-axis Top Hat jet core, we find $\theta_v = 0.114\pm0.004$ rad, $\log(E_0) = 54.51\pm0.03$, $\log(n_0)=1.11\pm0.05$, $p=2.023^{+0.005}_{-0.007}$, $\log(\epsilon_e)=-0.68\pm0.10$, and $\log(\epsilon_B) = -1.50^{+0.08}_{-0.07}$, see also Table \ref{table:fit_res} and the corner plot in Fig.\ref{Fig:corner_off-axis}. The total beaming-corrected energy is $1.5\times10^{51}$ erg. These parameters agree with studies of GRBs \citep{Aksulu2022}. The jet's main contribution is at high frequencies. In Table \ref{table:fit_res} we report the value of $A_{*}$ instead of the reference density $n_0$, to compare it with the refreshed shocks model. We find that the energetics of the two jet components (wings and core) is $\sim$5 orders of magnitude apart, with the off-axis core having an energetics in agreement with long GRBs \citep{Aksulu2022}. The other parameters, $\epsilon_e$, $\epsilon_B$ are in agreement within the two components, even if they are not necessarily expected to be equal. Instead, we expect similar values for $A_*$, which are in agreement within 2 $\sigma$.

The late-time optical data points (t $>20$ d) suggest a flattening in the light curve, which cannot be explained by the off-axis jet (dotted line). A natural way to account for it is by introducing a SN component, already associated with other XRFs \citep[as examples]{Bersier2006, Pian_2006}. In fact, adding the SN1998bw (at $z=0.75$) contribution would perfectly fit the optical data after 20 days, as demonstrated in Fig.\ref{Fig:lightcurve_fit_all}. 

\citet{Busmann2025} discard a structured jet as explanation for this afterglow because of the too steep rise of the flux during the bump. However, even if our modeling confirms that the fast rise of the bump is not perfectly taken into account, we show that this scenario well catches the overall flux behavior.

\subsubsection{Late-time radio afterglow}
\label{subsec:cocoon}

As demonstrated in Section \ref{sec:phenom}, the radio afterglow after $\sim$10 d is described by a bell-shaped spectrum, which cannot explain the emission in the optical and X-rays. A self-absorbed radio spectrum characterizes also the emission expected from a SN Ic \citep{Corsi2016}. However, the EP241021a luminosity in the radio band exceeds by more than 1 order of magnitude the expected luminosity of a radio SN. 
Moreover, FBOTs also show self-absorbed radio spectra \citep{Bright2022, Ho2019}, but their luminosity at 9 GHz is $\sim10^{28} \rm erg \ s^{-1} Hz^{-1}$, nearly $\sim$2 orders of magnitude lower than EP241021a radio emission.

As mentioned in Section \ref{sec:phenom}, we interpret the radio emission to be produced by a spherical ejecta, so-called cocoon (Cocoon I), which can be either nonrelativistic or mildly relativistic $(\gamma \lesssim 2)$, in a constant density environment (see Section \ref{sec:phenom}).
We used the {\tt{afterglowpy}} cocoon model, in which we assumed that the ejecta is made by one single shell, with velocity $u = \beta\gamma$, total kinetic energy $E_{\rm tot}$, and mass
\begin{equation}
    M = \frac{E_{\rm tot}}{(\sqrt{u^2 + 1}-1)c^2},    
\end{equation}
where $c$ is the velocity of light.
The fitted parameters and the respective (all uniform) priors are the following: the velocity of the ejecta $u$ in [0.1, 2]; the total kinetic energy $E_{\rm tot}$ in [$10^{48}$, $10^{52}$]; the density of the environment $n_0$ in [$10^{-3}$, $10^4$]; the slope of the accelerated electron population $p$ in [2,4]; the fraction of energy in the accelerated electrons $\epsilon_e$ in [0.3, 0.001]; and the fraction of energy in the magnetic field $\epsilon_B$ in [0.3, 0.001]. The prior bounds are inspired by \citet{Hotokezaka2015}, whose discussion, however, is mainly based on binary neutron star mergers. For this reason, we expand the parameter bounds regarding the energetics, the density and the mass of the ejecta.

The results are reported in the fourth column of Table \ref{table:fit_res}, while the corner plot is in Fig.\ref{Fig:corner_cocoon-main}. 
The fit is represented in Fig.\ref{Fig:lightcurve_fit_all} with dashed lines. The model fits well the radio data after 10 days, while the optical and X-ray emission do not influence the observed fluxes (the expected optical light curve is represented by the dashed purple line in the Figure, while the X-ray is not represented and reaches a peak flux of $\sim 2\times 10^{-8}$ mJy).
The cocoon ejecta is mildly relativistic, in agreement with \citet{Shu2025}, with a $\beta=0.73\pm0.01$,  $\gamma=1.455^{+0.015}_{-0.014}$ and a mass of $4^{+2}_{-4}\times10^{-3}M_{\odot}$. 
The density of the environment is quite high, with a median value of $2.8\times10^3 \rm cm^{-3}$.
Ejecta with $\gamma \geq 1.5$ at 10 d is inferred also in \citet{Yadav2025}, consistent with our findings for the cocoon. However, from a broad-band fit, they then identify these ejecta as a relativistic jet (Lorentz factor of 30-50), able to reproduce the X-ray, optical and radio afterglow (not including the bump). However, we note that this interpretation would result in a synchrotron self-absorption frequency $\nu_{sa}<< 1$ GHz, which does not fit the steep break that we find in the radio spectrum at $\nu\sim4$ GHz. Moreover, this model would not explain the early flat X-ray light curve exhibited by the EP-FXT data presented here (see Fig.\ref{Fig:lightcurve_fit_all} and Section \ref{subsec:early-time-ag}).

\subsubsection{ALMA component}
\label{subsec:cocoon_ALMA}

In the radio spectrum at 70 days, there is clearly a second component (Cocoon II, see Fig.\ref{Fig:radio_spectrum_PL}), peaking at $\sim$50 GHz and with a spectral shape again consistent with synchrotron self-absorption.
We fit this emission with a second cocoon; the priors and parameters are the same as reported in Section \ref{subsec:cocoon}. The results for the fit are written in Table \ref{table:fit_res}, last column. This cocoon is slower than the cocoon described in Section \ref{subsec:cocoon}, with Lorentz factor of $\gamma=1.056^{+0.002}_{-0.002}$, $\beta=0.32^{+0.01}_{-0.01}$ and a mass of $9^{+1}_{-2}\times10^{-3} \rm M_{\odot}$. The density of the environment is in agreement with the cocoon that produced the main radio afterglow, suggesting the same environment. 

We focused on a scenario that envisages a complex jet structure and associated cocoon.
However, we note that a double-peaked synchrotron spectrum can also emerge in a high-density medium, where the condition $\nu_c < \nu_a$ holds \citep{Gao2013}. Under these circumstances, a thermal component arises, besides the broken power-law synchrotron \citep{Ressler_2017}. This leads to a two-component (thermal and nonthermal) spectrum, with a peak at $\nu_a$, originating from the thermal component of the electron population, and a second peak at $\nu_m$. In this scenario, however, the synchrotron emission peaking at $\nu_m$ is expected to be broader than what we observe, with a spectral slope of 1/3 before the peak, which is inconsistent with the spectrum of EP241021a.

Another alternative explanation is that the second spectral peak arises from the reverse shock \citep{Kobayashi2004, Gao2013}, still under the regime $\nu_c < \nu_a$ with $\nu_a > \max(\nu_m, \nu_c)$. However, the detection of a refreshed shock emission at such late times appears highly improbable \citep{Laskar2018, Laskar2019}.
Finally, we find that in our cocoon models $\nu_c$ is much larger than $\nu_a$, contrary to the condition required by these scenarios.

\section{Discussion}
\label{sec:discussion}

   \begin{figure}
   \centering
   \includegraphics[width=0.9\hsize]{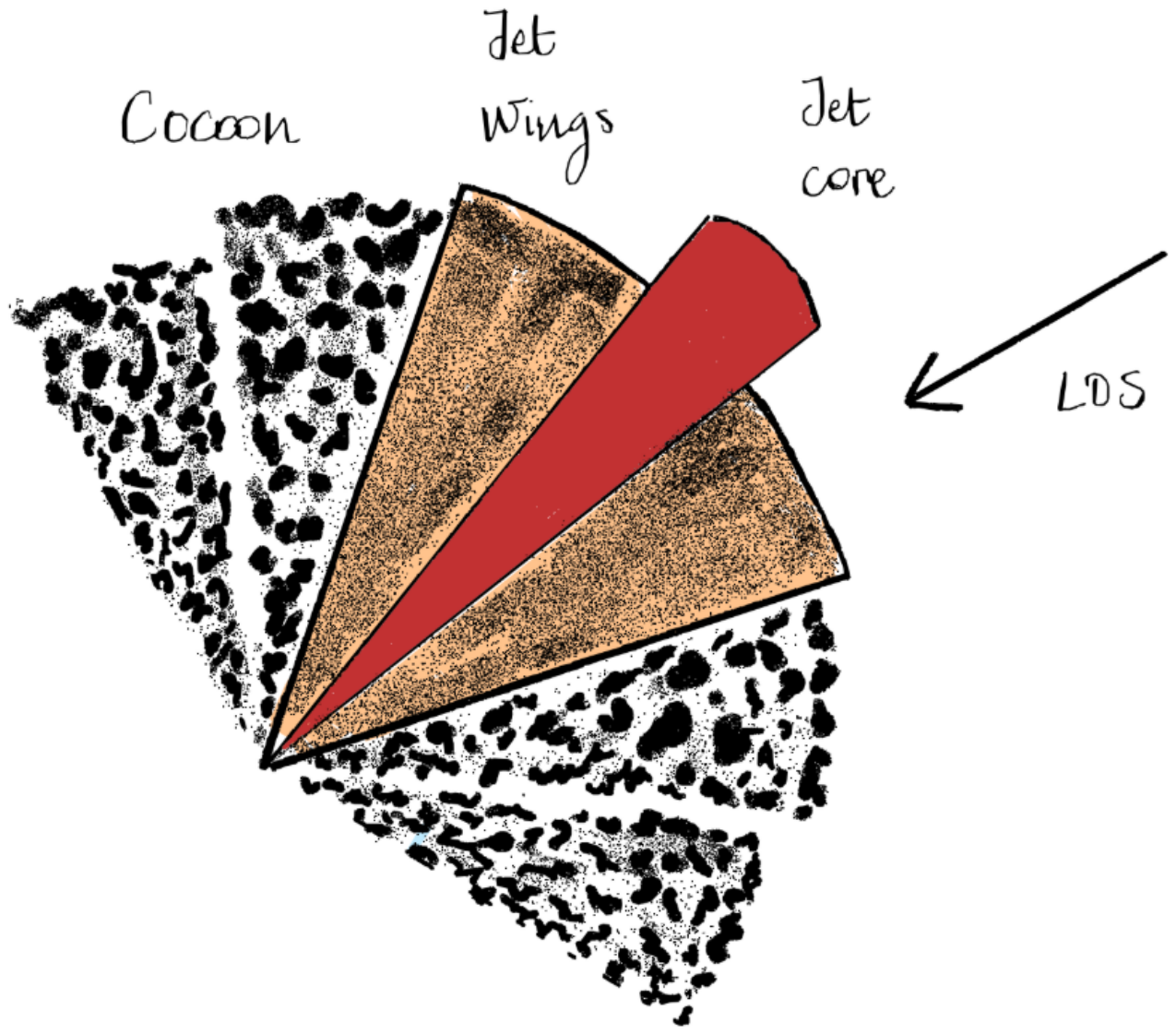}
      \caption{Sketch of the proposed scenario for EP241021a. At small polar angles, a relativistic jet with a top-hat core and wide wings is produced, while at large polar angles, the jet is surrounded by a structured cocoon (with a mildly relativistic and nonrelativistic component). The observer's line of sight lies within the wings of the jet. }
         \label{Fig:sketch}
   \end{figure}

\begin{figure}
   \centering
   \includegraphics[width=\hsize]{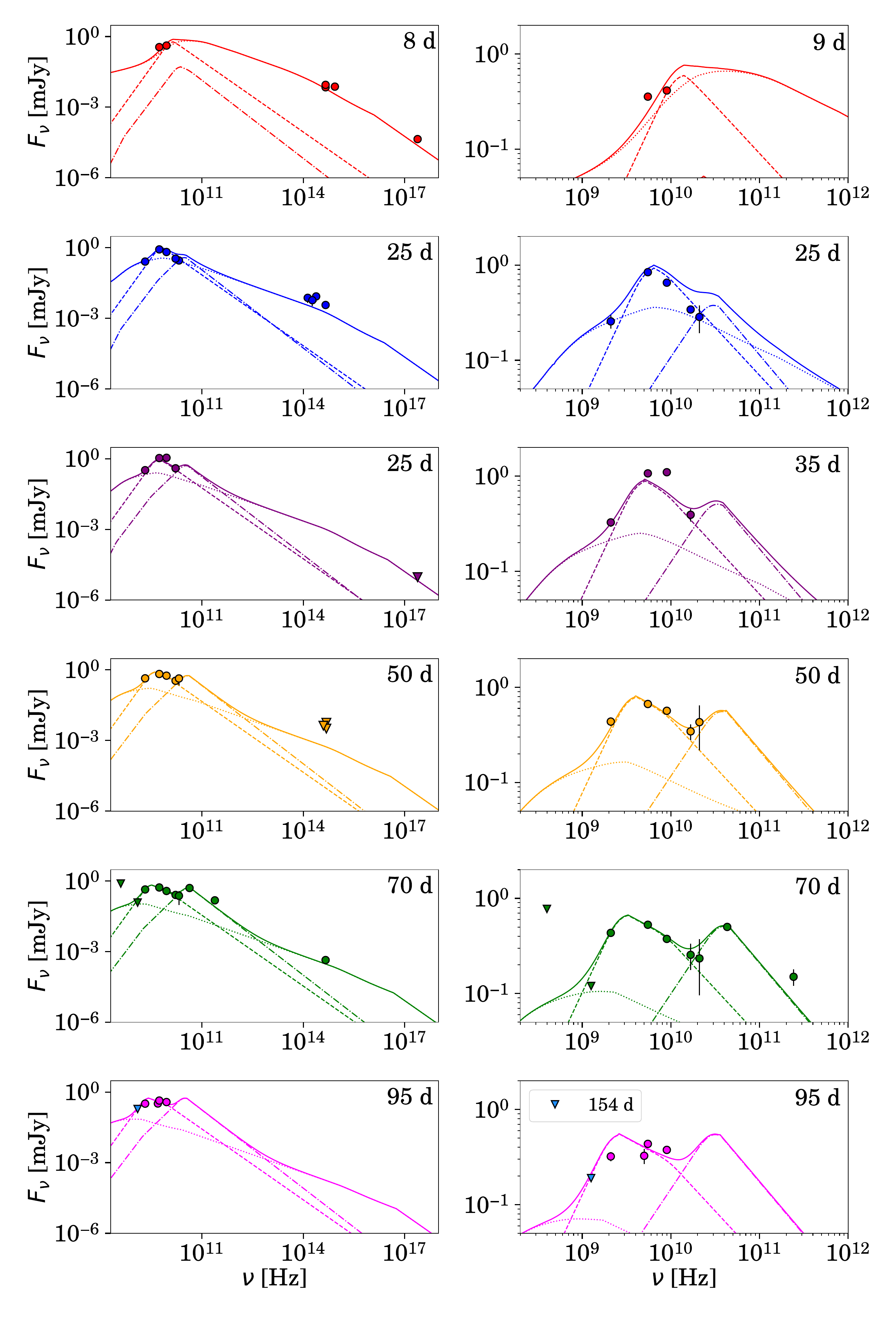}
      \caption{EP241021a spectra at different times, up to 70 days. The broadband spectrum is represented in the first column, while a zoom of the radio spectrum is represented in the second column. The observation date is shown in the top right of each plot. The solid line represents the sum of all the components: the relativistic jet producing the optical and X-ray emission (in dotted line), the main cocoon emission producing the radio data (dashed lines), and the cocoon producing the ALMA data (dot-dashed lines). In the last two panels, the blue upper limit refers to the u-GMRT observation at $\sim$154 days, while the other data points refer to those taken at $\sim95$ days. }
         \label{Fig:spectrum_fit_all}
   \end{figure}

The physical modeling of EP241021a suggests that several components come into play, the complete spectral fit is represented in Fig.\ref{Fig:spectrum_fit_all}. The system at small polar angles is composed of one structured jet with an energetic top-hat core and external wider and low-Lorentz-factor wings, see a sketch in Fig.\ref{Fig:sketch}. Our line of sight is within the wings, but outside the collimated core. This results in the emission being dominated first by the wings (slow decay in optical and radio band), and later by the collimated core (bump in the optical and X-ray), once it enters our line of sight. The prompt emission in the soft X-rays is likely originated in the wide baryon-loaded wings, as it is also theorized in \citealt{Busmann2025}. We cannot rule out the presence of a SN; in fact, it would provide a natural explanation for the shallow decline of the optical light curve at late times \citep[see also][]{Busmann2025}.
The geometry of this system, with on-axis wide and less energetic wings and an off-axis core, could explain why previous analyses of the XRF afterglow population \citep{DAlessio_2006, Gendre2007} found no compelling evidence for XRFs to originate from an off-axis jet. As in the EP241021a case, a less energetic and wider component could dominate the emission at early times, masking the typical rising light curve of off-axis GRBs.

The radio emission at $\nu<20$ GHz is produced by a spherical mildly relativistic ejecta, the cocoon, which is located at large polar angles (see Fig.\ref{Fig:sketch}). Possibly, this is a stratified cocoon with two different velocities, explaining the second spectral peak found at 70 days and $\nu>20$ GHz. The spectrum is self-absorbed, with a peak frequency ($\nu_{sa}$) of $\sim5$ GHz. Such a high $\nu_{sa}$ suggests a high density, of $10^3 \rm cm^{-3}$, which is in contrast with the lower density found by the jets.

Therefore, this system suggests an anisotropic medium. Taking as a reference a radius of $10^{17}$ cm, the cocoon expands in a constant density medium with $\sim10^{3} \rm cm^{-3}$, while the two jets suggest a much lower $1-10 \rm cm^{-3}$. The densities from the two jet components are in agreement with each other, as the jet wings have a wide constraint on the density, see Table\ref{table:fit_res}. 
These differences can be explained by an anisotropic medium with higher densities at higher polar angles, shaped by the wind progenitor. This scenario agrees with the evidence of anisotropic winds in WR stars associated with their large rotational velocity \citep{Shenar2019, Callingham2019}. The slower, denser wind is ejected along the equatorial plane, where the cocoon is expanding, likely encountering a constant density termination shock produced by the denser wind interacting with the ISM. On the other hand, the faster, low-density component of the wind is ejected along the rotation axis, hence in the direction encountered by the relativistic jet.

After the collapse of a massive star, 3D hydrodynamical simulations predict the presence of a cocoon --with an inner mildly relativistic part and an outer nonrelativistic part -- alongside a structured jet, where the wings represent a transition area between the energetic and relativistic core of the jet and the cocoon \citep{Gottlieb2020, Harrison2018, Ito2015, Mizuta2013}.
In fact, such a jet-cocoon system has been proposed to explain other soft X-ray transients discovered by EP.
EP240414a \citep{Hamidani2025b, Zheng2025} shows optical re-brightening, interpreted with a nonthermal component: either an off-axis jet \citep{Zheng2025} or a mildly relativistic cocoon \citep{Hamidani2025a}, followed by SN emission. The luminosity of the radio afterglow of this transient, produced by the mildly relativistic cocoon, is located between low-luminosity GRBs and cosmological GRBs, in agreement with our findings for EP241021a.
Furthermore, a nonrelativistic cocoon is identified through a thermal spectrum in the optical at $t<1.2$ d. 
In the case of EP241021a, observations start only later than $\sim$1 day, and thermal emission from the nonrelativistic cocoon identified in the high frequency radio band cannot be ruled out. 
EP240108a \citep{Li_EP250108a, Eylesferris2025, rastinejad2025} shows a shallow decay in the optical afterglow, before showing re-brightening due to a SN at $\sim10$ d \citep{Eylesferris2025, rastinejad2025}, but there were no detections at radio and X-ray wavelengths. The early decaying phase is interpreted as the afterglow from a jet with a moderate Lorentz factor or as shock-cooling emission from a cocoon \citep{Li_EP250108a}.

Also in past GRBs, there is evidence of systems with a jet and a cocoon, such as in GRB 130925A, where X-ray thermal emission is detected \citep{Piro2014}, and GRB 160623A, where a cocoon produces the radio afterglow \citep{Chen2020}. In general, multicomponent jets were found also in XRF 030723 \citep{Huang2004}, GRB030329 \citep{Berger2003, Resmi2005}, GRB 080413B \citep{Filgas2011}, GRB 221009A \citep{Sato2023, Laskar2023, OConnor2023}).

\section{Conclusions}
\label{sec:conclusions}

We presented a comprehensive multiwavelength analysis of EP241021a, a soft X-ray transient detected by the Einstein Probe, interpreting its properties within the framework of XRFs. The prompt emission characteristics, along with the isotropic distribution and soft spectral peak, support the classification of EP241021a as an XRF.

Our observations reveal a complex, multicomponent afterglow, consistent with theoretical models involving structured jets and cocoon emissions. The optical and X-ray light curves at early times are best explained by wide-angled, low-Lorentz-factor structured wings, followed by the delayed emergence of the off-axis jet core. The radio afterglow is well modeled by two distinct cocoon components—mildly (at $\nu\leq20$ GHz) and nonrelativistic ($\nu>20$ GHz)—implying a stratified outflow geometry.

Furthermore, our findings suggest significant asymmetry in the circumstellar medium, with a low-density region along the jet axis and denser material at wider angles, consistent with the wind structure of WR progenitors. This supports the theory that XRFs represent a softer and/or off-axis manifestation of long GRBs shaped by both viewing angles and explosion dynamics.

We find that the EP241021a emission aligns with a collapsar scenario. However, due to its complexity, other possible interpretations can be put forward, such as a complex tidal disruption event, proposed in \citet{Shu2025}, or a catastrophic collapse or merger of a compact star system, which led to the formation of a millisecond magnetar, in \citet{Wu2025}.

Thanks to its sensitivity and ability to detect soft X-ray energies, EP has revealed a landscape that remained hidden for years. As the number of XRFs continues to increase, it is possible that some or even all of these phenomena can eventually be explained by these multiple components interacting over various timescales.
Crucially, studying these transients requires coordinated observations across multiple facilities and wavelengths. As demonstrated in this event, each energy band provides a piece of the story, and only by combining them can we gain a complete understanding of the system producing the observed transient.

\begin{acknowledgements}
The authors thank the anonymous referee for the useful comments.
The authors also thank Bing Zhang for stimulating discussion.
This work is based on data obtained with the Einstein Probe, a space mission supported by the Strategic Priority Program on Space Science of the Chinese Academy of Sciences, in collaboration with ESA, MPE and CNES (Grant No. XDA15310000, No. XDA15052100).
We acknowledge support by the European Union Horizon 2020 programme under the AHEAD2020 project (grant agreement number 871158). This work has been also supported by ASI (Italian Space Agency) through the Contract no. 2019-27-HH.0.\\
AJCT acknowledges support from the Spanish Ministry project PID2023-151905OB-I00 and Junta de Andaluc\'ia grant P20$\_$010168 and from the Severo Ochoa grant CEX2021-001131-S funded by MCIN/AEI/10.13039/501100011033. MCG acknowledges support from the Spanish Ministry project PID2023-149817OB-C31.\\
e-MERLIN is a National Facility operated by the University of Manchester at Jodrell Bank Observatory on behalf of STFC.  
This project has received funding from the European Union’s Horizon 2020 research and innovation programme under grant agreement No 101004719.\\
Based on observations collected at the Centro Astron\'omico Hispano en Andaluc\'ia (CAHA) at Calar Alto, proposal 24B-2.2-012, operated jointly by Junta de Andaluc\'ia and Consejo Superior de Investigaciones Cient\'ificas (IAA-CSIC). Also based on observations made with the Gran Telescopio Canarias (GTC), installed at the Spanish Observatorio del Roque de los Muchachos of the Instituto de Astrofísica de Canarias, on the island of La Palma.\\
The Australia Telescope Compact Array is part of the Australia Telescope National Facility (grid.421683.a), which is funded by the Australian Government for operation as a National Facility managed by CSIRO. We acknowledge the Gomeroi people as the traditional owners of the Observatory site.\\
We thank the staff of the GMRT that made these observations possible. GMRT is run by the National Centre for Radio Astrophysics of the Tata Institute of Fundamental Research.\\
This paper makes use of the following ALMA data: ADS/JAO.ALMA\#2024.A.00019.T. ALMA is a partnership of ESO (representing its member states), NSF (USA) and NINS (Japan), together with NRC (Canada), NSTC and ASIAA (Taiwan), and KASI (Republic of Korea), in cooperation with the Republic of Chile. The Joint ALMA Observatory is operated by ESO, AUI/NRAO and NAOJ.\\

\end{acknowledgements}

\bibliographystyle{aa} 
\bibliography{bib} 

\FloatBarrier

\begin{appendix}
\onecolumn

\section{Observations}

\begin{table}[h!]
\caption{EP-FXT observation log and flux measurements.}             
\label{table:xray}      
\centering          
\begin{tabular}{c c c c c c}
\hline\hline       
T-T$_{0}$ & ObsID & Effective exposure & Count rate & Unabsorbed flux (0.5-10 keV)$^{(a)}$ & Flux density (1 keV) \\
 $[\rm d]$ &  & [ks] & [10$^{-3}$ cts s$^{-1}$] & [10$^{-13}$ erg cm$^{-2}$ s$^{-1}$] & [10$^{-6}$ mJy] \\
\hline                    
   1.54 & 06800000167 & 2.90 & $10.3\pm2.1$ & $3.18\pm0.64$ & $37.91\pm7.64$ \\  
   2.57 & 06800000168 & 5.90 & $9.5\pm1.4$ & $2.94\pm0.43$ & $35.04\pm5.15$ \\
   3.64 & 06800000170 & 5.90 & $8.7\pm1.3$ & $2.69\pm0.42$ & $32.12\pm4.97$ \\
   4.62 & 06800000173 & 8.30 & $7.3\pm1.1$ & $2.27\pm0.33$ & $27.05\pm3.94$ \\
   5.61 & 06800000176 & 0.85 & $3.0\pm2.7$ & $<2.99$ & $<35.7^{(c)}$ \\
   6.45 & 06800000181 & 5.90 & $8.3\pm1.3$ & $2.55\pm0.39$ & $30.45\pm4.76$ \\
   8.19 & 06800000186 & 5.30 & $11.2\pm1.6$ & $3.45\pm0.48$ & $41.11\pm5.73$ \\
   9.19 & 06800000192 & 1.40 & $4.9\pm2.4$ & $<3.32$ & $<39.6^{(c)}$ \\
   12.42 & 06800000198 & 2.00 & $5.7\pm2.1$ & $1.78\pm0.65$ & $21.18\pm7.73$ \\
   14.14 & 06800000202 & 5.80 & $4.1\pm1.1$ & $1.25\pm0.35$ & $14.93\pm4.14$ \\
   17.33 & 06800000211 & 4.60 & $4.3\pm1.2$ & $1.34\pm0.36$ & $16.01\pm4.34$ \\
   30.50 & 06800000250 & 6.00 & $2.4\pm0.9$ & $<0.73$ & $<8.70^{(c)}$ \\
   40.67 & 06800000269 & 8.90 & $1.3\pm0.6$ & $<0.40$ & $<4.76^{(c)}$ \\
   52.46 & 06800000291 & 0.33 & $4.1\pm4.5$ & $<6.2$ & $<73.9^{(c)}$ \\
   79.32 & 06800000356 & 8.80 & $1.2\pm0.9^{(b)}$ & $<0.39$ & $<4.60^{(c)}$ \\
\hline
\end{tabular}
\tablefoot{
    \tablefoottext{a}{The counts-to-flux factor is $3.1\times10^{-11}$, derived from a Galactically absorbed ($N_H = 5\times10^{20} \mathrm{cm}^{-2}$) power law of index $1.8$} \\
    \tablefoottext{b}{Count rate from FXT-A only} \\
    \tablefoottext{c}{3${\sigma}$ upper limit}
    }
\end{table}

\begin{table}[h!]
\caption{Optical observations of EP241021a.}             
\label{table:optical}      
\centering          
\begin{tabular}{c c c c c}
\hline\hline       
Time &  Telescope/ & Filter & Exposure & AB magnitude \\
 $[\rm d]$ & Instrument &  & [s] & \\
\hline                    
7.05 & LBT & r	& 600 & $21.78\pm0.02$ \\
7.05 & LBT & z & 600 & $21.25\pm0.03$ \\
23.3 & GTC/EMIR & J & 350 & $21.58\pm0.37$ \\
23.3 & GTC/EMIR & H & 462 & $21.98\pm0.39$ \\
23.3 & GTC/EMIR & Ks & 588 & $21.73\pm0.36$ \\
47.8 & CAHA/CAFOS & g$^\prime$ & 840 & $>22.94$ \\
47.8 & CAHA/CAFOS & r$^\prime$ & 900 & $>22.64$ \\
47.8 & CAHA/CAFOS & i$^\prime$ & 900 & $>22.33$ \\
105.63 & GTC/EMIR & J & 350 & $>23.53$ \\
105.63 & GTC/EMIR & H & 462 & $>23.41$ \\
105.63 & GTC/EMIR & Ks & 588 & $>23.83$ \\
\hline                  
\end{tabular}
\tablefoot{ The AB magnitudes are not corrected for the foreground Galactic extinction.}
\end{table}

\twocolumn

\begin{table}
\caption{Radio observations of EP241021a.}             
\label{table:radio} 
\centering          
\begin{tabular}{c c c c} 
\hline\hline       
Time & Telescope & Frequency & Flux \\
 $[\rm d]$ &  & [GHz] & [mJy] \\
\hline                    
3.5 & e-MERLIN & 5.0 & $<$0.2 \\
8.4 & ATCA & 5.5 & 0.356$\pm$0.018 \\
8.4 & ATCA & 9 & 0.413$\pm$0.021 \\
19.5 & e-MERLIN & 	5 & 0.614$\pm$0.164 \\
21.1 & ATCA & 	2.1 & 0.256$\pm$0.042 \\
25.1 & ATCA	 & 16.7	 & 0.341$\pm$0.029 \\
25.1 & ATCA & 	21 & 	0.285$\pm$0.092 \\
25.1 & ATCA	 & 5.5 & 	0.842$\pm$0.044 \\
25.1 & ATCA	 & 9.0 & 	0.653$\pm$0.034 \\
31.6 & ATCA	 & 2.1 & 	0.327$\pm$0.033 \\
31.6 & ATCA & 	5.50 & 	1.07$\pm$0.09 \\
31.6 & ATCA	 & 9.0 & 	1.1$\pm$0.07 \\
34.1 & ATCA	 & 16.7 & 	0.393$\pm$0.061 \\
34.1 & ATCA	 & 21 & 	0.317$\pm$0.327 \\
47.1 & ATCA	 & 5.5 & 	0.667$\pm$0.036 \\
47.1 & ATCA	 & 9.0 & 	0.566$\pm$0.031 \\
47.1 & ATCA	 & 16.7 & 	0.344$\pm$0.065 \\
47.1 & ATCA	 & 21 & 	0.43$\pm$0.213 \\
48.1 & ATCA	 & 2.1 & 	0.436$\pm$0.044 \\
68.7 & ALMA	 & 40 & 	0.501$\pm$0.009 \\
68.9 & ALMA	 & 233 & 	0.15$\pm$0.03 \\
70.0 & ATCA & 2.1 & 	0.434$\pm$0.022 \\
70.1 & ATCA	 & 5.5 & 	0.528$\pm$0.033 \\
70.1 & ATCA	 & 9.0 & 	0.375$\pm$0.023 \\
70.2 & ATCA	 & 16.7 & 	0.255$\pm$0.078 \\
70.2 & ATCA	 & 21 & 	0.234$\pm$0.138 \\
70.3 & uGMRT	 & 0.4 & 	$<$0.775	 \\
70.5 & uGMRT	 & 1.26 & 	$<$0.120	 \\
92.3 & e-MERLIN	 & 5.0 & 0.326$\pm$0.06 \\
96.0 & ATCA & 2.1 & 	0.321$\pm$0.036 \\
96.0 & ATCA & 5.5 & 	0.435$\pm$0.032 \\
96.0 & ATCA & 9.0 & 	0.376$\pm$0.023 \\
153.5 & uGMRT & 1.26 & $<$0.114	 \\
153.8 & e-MERLIN & 5.0 & 	0.15$\pm$0.031 \\
\hline                  
\end{tabular}
\end{table}

\section{Corner plots and additional figures}

   \begin{figure}[h!]
   \centering
   \includegraphics[width=\hsize]{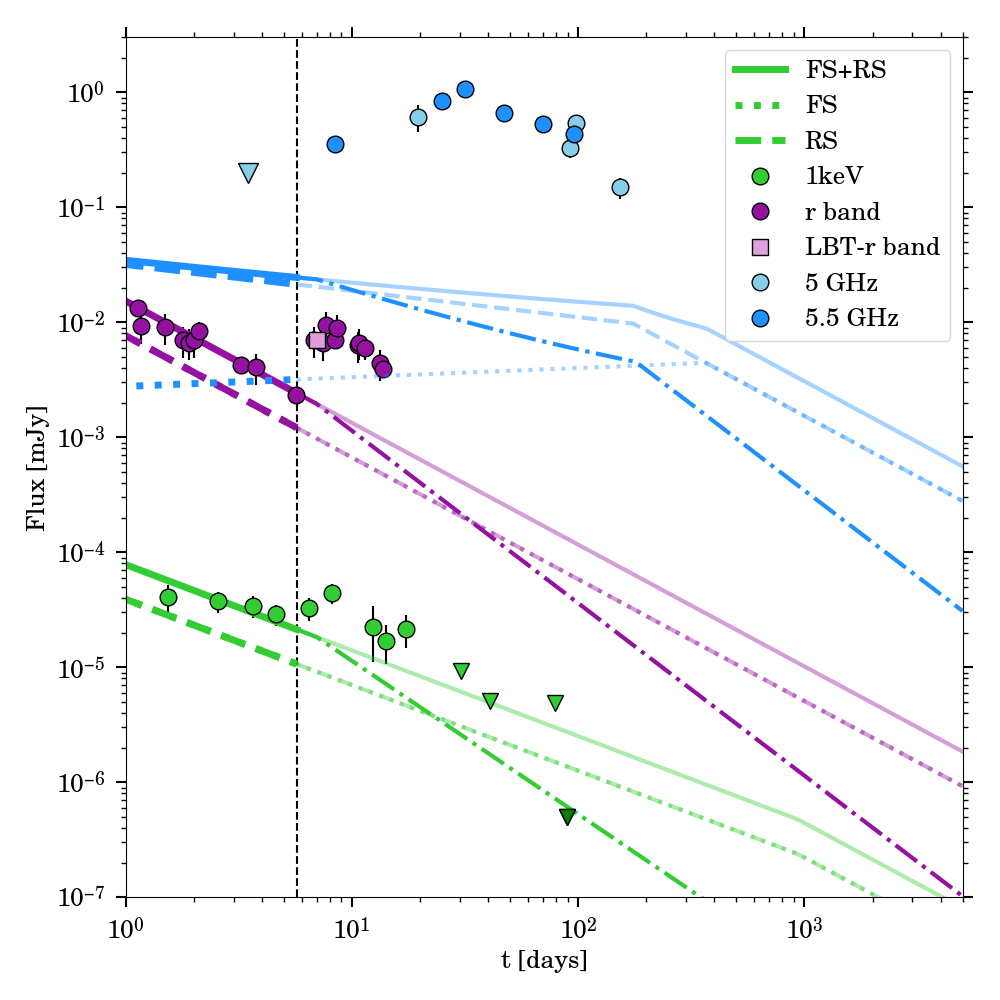}
      \caption{EP241021a broadband light curve. The dashed line represents the contribution of the RS, the dotted line represents the contribution of the FS, while the solid line represents the sum of FS and RS. The FS and RS are superimposed in the optical and X-ray bands. The data points before the vertical dashed line are fit, and the model following that is an interpolation, assuming that the refreshing phase continues up to $5\times10^3$ days. The dot-dashed line represents the model interpolation when the refreshing phase ends at 10 days. The dark green X-ray upper limit at 89 d is derived from \citealt{Shu2025}.}
         \label{Fig:lightcurve_refreshed}
   \end{figure}

\FloatBarrier

\clearpage

\onecolumn

   \begin{figure}
       \centering
   \includegraphics[width=0.8\hsize]{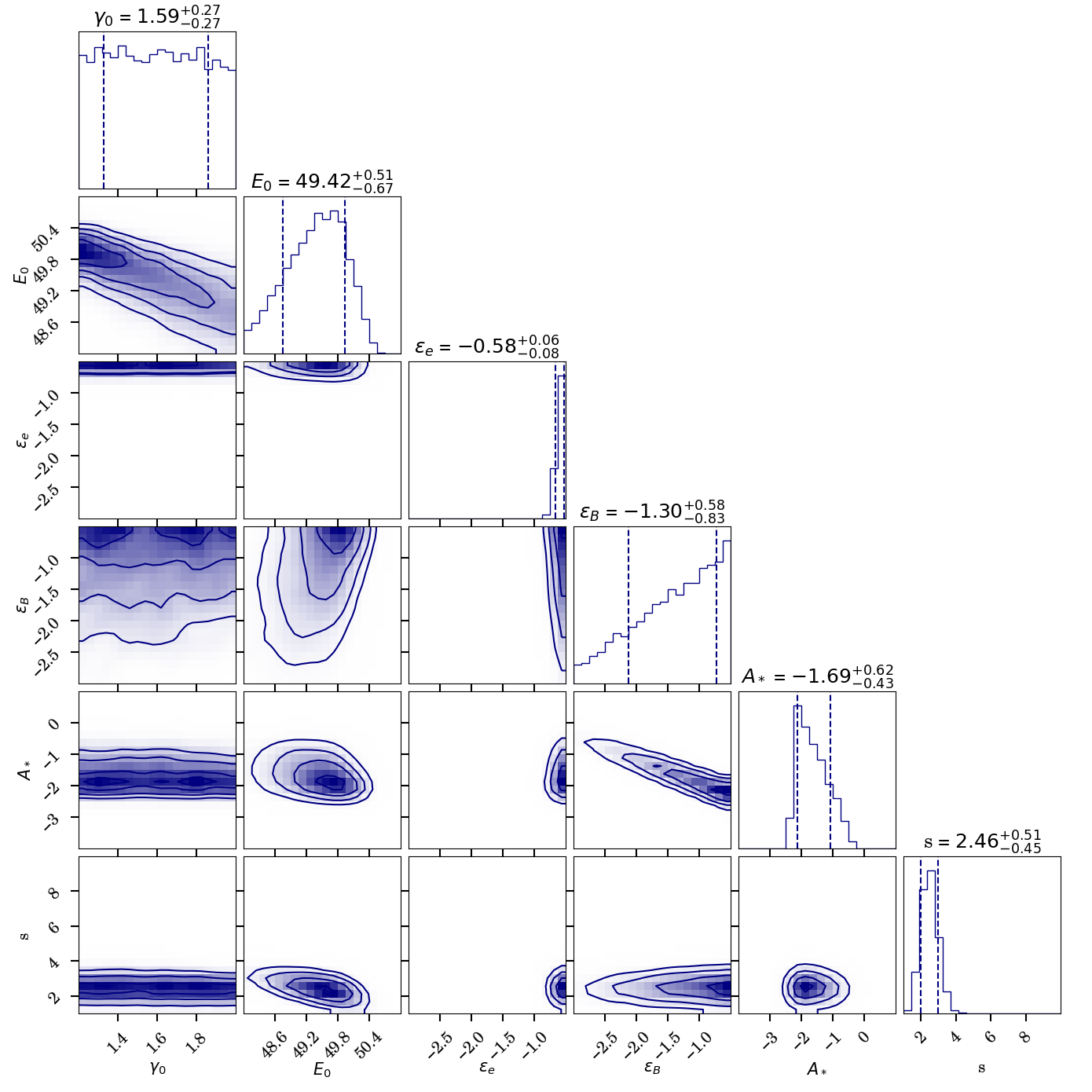}
      \caption{Corner plot of the early-time ($t\leq 6$d) light curve fit using the refreshed shocks model. The dashed lines in the 1D histograms represent the 16th and 84th percentiles. The parameter values are indicated as median, 16th, and 84th percentiles.}
         \label{Fig:corner_refreshed}
   \end{figure}

   \begin{figure}
       \centering
   \includegraphics[width=0.8\hsize]{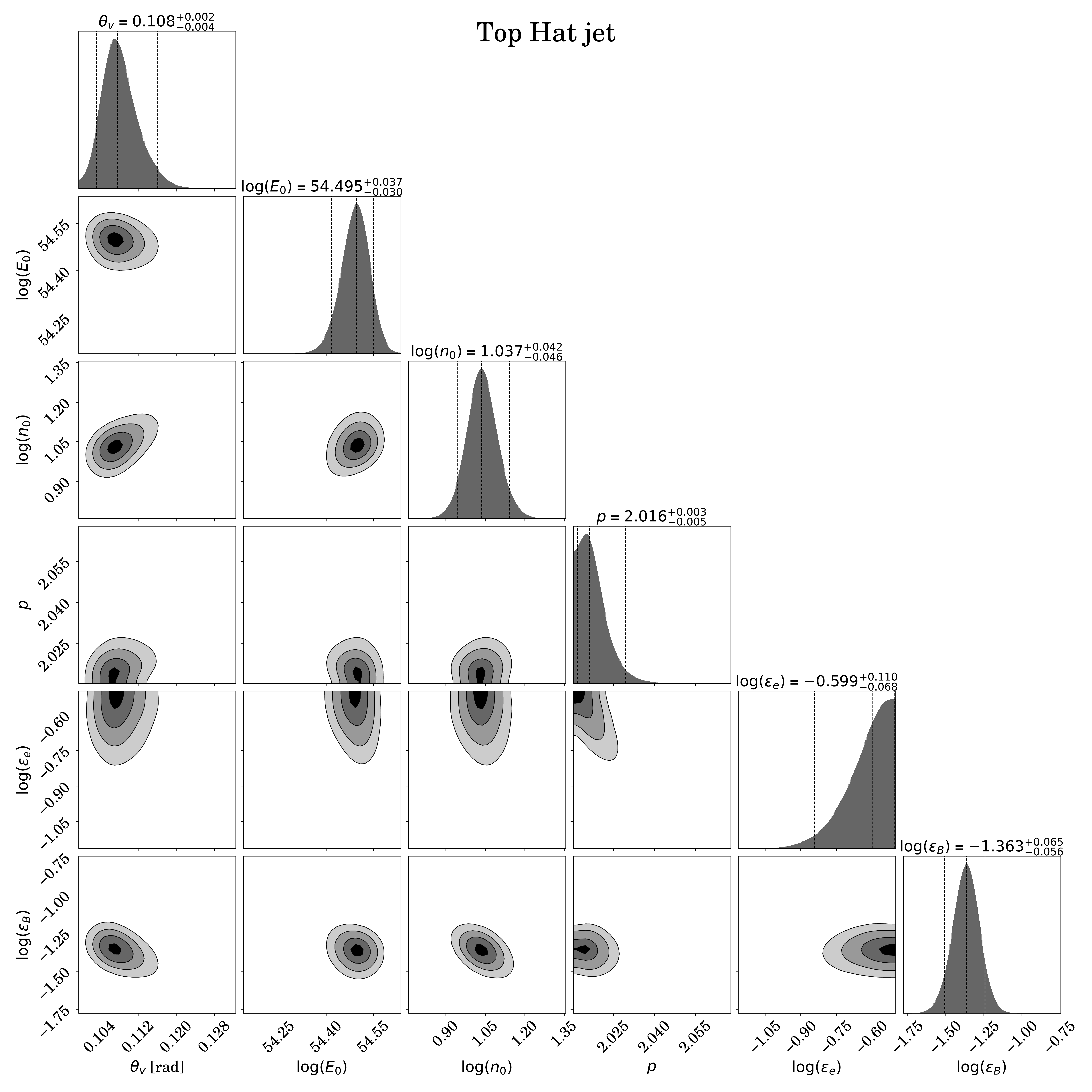}
      \caption{Corner plot of the off-axis top-hat jet. The dashed lines in the 1D histograms represent the 16th and 84th percentiles. The parameter values are indicated as median, 16th, 50th, and 84th percentiles.}
         \label{Fig:corner_off-axis}
   \end{figure}

    \begin{figure}
       \centering
   \includegraphics[width=0.8\hsize]{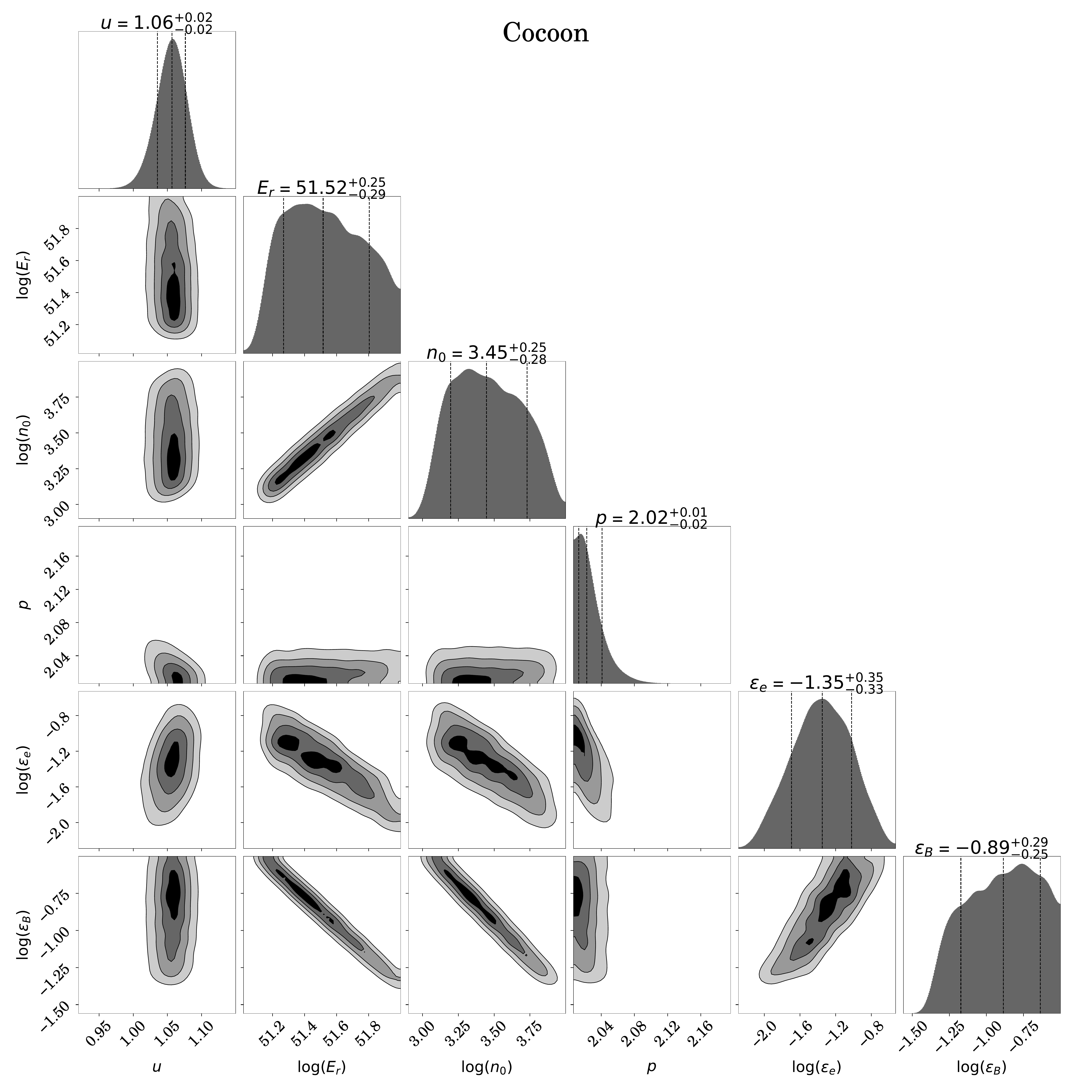}
      \caption{Corner plot of the cocoon fitting the radio data. The dashed lines in the 1D histograms represent the 16th and 84th percentiles. The parameter values are indicated as median, 16th, 50th, and 84th percentiles.}
         \label{Fig:corner_cocoon-main}
   \end{figure}

\end{appendix}

\end{document}